\documentclass{aa}  

\usepackage{graphicx}
\usepackage{multirow}
\usepackage{txfonts}
\usepackage{amsfonts}
\usepackage[colorlinks=True,citecolor=blue,linkcolor=blue]{hyperref}
\usepackage{xcolor}

\begin{document}

   \title{A case study of early galaxy cluster with the \emph{Athena} X-IFU}


   \author{F. Castellani\inst{1}
          \and
          N. Clerc\inst{1}
          \and
          E. Pointecouteau\inst{1}
          \and
          Y.M. Bahé\inst{2,3}
          \and
          J. Schaye\inst{3}
          \and
          F. Pajot\inst{1}}

   \institute{IRAP, Université de Toulouse, CNRS, UPS, CNES, Toulouse, France\\
              \email{nicolas.clerc@irap.omp.eu}
         \and
             Institute of Physics, Laboratory of Astrophysics, Ecole Polytechnique Fédérale de Lausanne (EPFL), Observatoire de Sauverny, 1290 Versoix, Switzerland
        \and
             Leiden Observatory, Leiden University, P.O. Box 9513, 2300 RA, Leiden, The Netherlands}

   \date{Received - ; accepted -}

 
  \abstract
{Observations of the hot gas in distant clusters of galaxies, though challenging, are key to understand the role of intense galaxy activity, super-massive black hole feedback and chemical enrichment in the process of massive halos assembly.}
   {We assess the feasibility to retrieve, using X-ray hyperspectral data only, the thermodymamical hot gas properties and chemical abundances of a $z=2$ galaxy cluster of mass $M_{500} = 7 \times 10^{13}$\,M$_{\odot}$, extracted from the Hydrangea hydrodynamical simulation.}
   {We create mock X-ray observations of the future X-ray Integral Field Unit (X-IFU) onboard the \emph{Athena} mission. By forward-modeling the measured $0.4-1$\,keV surface brightness, the projected gas temperature and abundance profiles, we reconstruct the three-dimensional distribution for the gas density, pressure, temperature and entropy.}
   {Thanks to its large field-of-view, high throughput and exquisite spectral resolution, one X-IFU exposure lasting 100\,ks enables reconstructing density and pressure profiles with $20$\% precision out to a characteristic radius of $R_{500}$, accounting for each quantity’s intrinsic dispersion in the Hydrangea simulations.
    Reconstruction of abundance profiles requires both higher signal-to-noise ratios and specific binning schemes. We assess the enhancement brought by longer exposures and by observing the same object at later evolutionary stages (at $z=1$ and $1.5$).}
   {Our analysis highlights the importance of scatter in the radially binned gas properties, which induces significant effects on the observed projected quantities. The fidelity of the reconstruction of gas profiles is sensitive to the degree of gas components mixing along the line-of-sight. Future analyses should aim at involving dedicated hyper-spectral models and fitting methods that are able to grasp the complexity of such three-dimensional, multi-phase, diffuse gas structures.}


   \maketitle
%
\section{Introduction}

The formation epoch of groups and clusters of galaxies ranges from $z\sim 1-3$ when star formation in galaxies and super-massive black holes (SMBH) are at the peak of their activities. The gas trapped in the forming massive potential wells of these structures heats up under the dual effect of gravity and feedback from star formation and SMBH \citep[e.g.,][]{kra12,vog14, Schaye2015,bah17,Schaye2023}. 
The questions of how this gas is accreted by galaxies and how it feeds their star formation and SMBH activity are still open. The same goes for the timescale within which galaxies evolve into the massive ellipticals that form the red sequence under the joint influence of their dense environment and the quenching of star formation by active galactic nucleus (AGN) feedback \citep[e.g.,][]{beh13,lea12, eck21,opp21}. This phase of violent and intense astrophysical activity injects large amounts of energy, gas and metals into the forming intra-cluster medium, shaping its thermal and chemical properties \citep[e.g.,][]{mcn07,bif17, mer18}. As such, these processes imprint the statistical (scaling and structural) properties of the population of groups and clusters of galaxies \citep[][for a recent review]{lov22}. Constraining the properties and the evolution of hot gas in these massive halos out to their epoch of formation is  an efficient way to understand the aforementioned assembly and evolution of the largest gravitationally bound halos in the Universe.

By construction in a hierarchical scheme of structure formation, lower mass halos ($M_{500}<10^{14}$~M$_\odot$) constitute the vast majority of the population of groups of galaxies \cite[e.g.,][]{tin08}. They are observed in numbers in large surveys, especially in optical surveys \citep[e.g.,][]{lam20,wer23}, and they dominate the population of simulated halos in numerical simulations. The self-similar process of structure formation predicts their properties to be down-scaled versions of massive galaxy clusters. 
Though these predictions seem to agree with some of their statistical properties \citep[e.g., the mass-temperature relation,][]{bab23},  many observations have shown that most actually depart from the scaling and structure behaviour of their more massive siblings \citep[e.g.,][]{pon03,min11,Stott2012,san13,lov15,lov21}. Their shallower potential is more prone to the impact of AGN feedback in terms of intra-group (IGM) gas heating, but also in the way it impacts the IGM gas distribution and its depletion from these less deep gravitational potential wells \citep[e.g.,][]{eck21}. 
The physics governing the gas content of galaxy groups and clusters is a current true challenge for numerical simulations, the current predictions of which largely vary from one work to the other \citep{opp21}.
    
With their smaller masses and thus shallower potential well with respected to massive galaxy clusters, the observation of the X-ray emitting hot gas is harder for groups than for clusters. To date several groups or samples of groups of galaxies  have been studied at X-ray wavelengths, mainly in the local Universe \citep[e.g.,][]{lov22}. Reaching such objects out to larger redshifts to investigate their gas content and its evolution remains very challenging  for the current generation of X-ray telescopes. 
The next generation of X-ray observatories shall combine a large collective area with high resolution spectro-imaging capabilities. The X-IFU instrument \citep{barret2023} on board the future European X-ray observatory \emph{Athena} \citep{bar17} should enable studies of groups with masses of a few $10^{13}$~M$_\odot$ out to $z \approx 2$. The understanding of the assembly of structure, and more specifically of massives halos, is a key objective of the Hot and Energetic Universe science theme that the \emph{Athena} mission will implement \citep{nan13}. 

In the wake of other previous feasibility studies addressing the science cases of chemical enrichment in groups and galaxy clusters \citep{Cucchetti2018, mer20}, bulk and turbulent motion in the intra-cluster gas \citep{ron18,cle19,Cuchetti2019}, and the warm hot intergalactic medium \citep{wal20,wij20,wij22}, we address in this work the issue of the observation of distant galaxy groups and clusters in order to characterise their physical  properties out to the epoch of their formation with the \emph{Athena} X-IFU instrument. Following the Athena Mock Observing Plan\footnote{Version v4.3 issued on 8$^{th}$ of September 2020 by the ESA Athena Science Study Team, the science objective concerning the evolution of thermodynamical properties of groups and clusters of galaxies from $z>0.5$ and out to $z\sim 2$ will be achieved with X-IFU observations. We therefore focused in this study on this specific instrument only.} We present realistic mock observations with the X-IFU instrument of one simulated galaxy cluster extracted from the cosmological hydrodynamic simulations Hydrangea  \citep{bah17}. 
The paper is organised as follows. We present the input of our simulations in Sec.~\ref{sect:simulations} and the mock X-IFU observations and processing in Sec.~\ref{sect:mock_observations}. In Sec.~\ref{s:mod}, we describe our Bayesian approach analysis. We present our results in  Sec.~\ref{sect:results} and discuss them in Sec.~\ref{sect:discussion}.

Throughout this paper, we made use of the cosmology setup adopted for the Hydrangea simulations, that is the \citet{pla14} cosmology with $h_{100}=0.6777$, $\Omega_m=0.307$, $\Omega_\Lambda=0.693$. At a redshift of $z = 2$, 1~arcsecond corresponds to a physical size of 8.6~kilo-parsec (kpc).

\section{A simulated cluster of galaxies at $z=2$}
\label{sect:simulations}

As input for our study, we used a simulated cluster of galaxies from the cosmological Hydrangea simulations \citep{bah17}.

\subsection{Hydrangea simulation}
\label{s:hyd}

Hydrangea is a suite of 24 cosmological zoom-in simulations of massive galaxy clusters (selected such that $M_{200} = 10^{14-15.4} M_{\odot}$ at $z=0$) that is part of the Cluster-{\small{\textsc{EAGLE}}} ({\small{\textsc{C-EAGLE}}}) project \citep{Barnes2017}. Hydrangea adopts the {\small{\textsc{EAGLE}}} \citep{Schaye2015,Crain2015} galaxy formation model but for zooms of regions taken from a larger volume (3200 cMpc)$^3$ than the original {\small{\textsc{EAGLE}}} parent volumes of $\leq$(100 cMpc)$^3$.

Hydrangea uses a modified version of the \textit{N}-Body Tree-PM smoothed particle hydrodynamics (SPH) code \textsc{gadget}-3 \citep{Springel2005}, with hydrodynamical updates by \cite{Schaye2015} and \cite{Schaller2015}. The subgrid physics of the code is based on that developed for OWLS \citep{Schaye2010}: it implements for instance radiative cooling and photoheating \citep{Wiersma2009a} for 11 chemical elements (H, He, C, N, O, Ne, Mg, Si, S, Ca, and Fe), hydrogen reionization and ionizing UV/X-ray background \citep{Haardt2001}, star formation rate of gas following \cite{Schaye2008}, stellar mass loss based on \cite{Wiersma2009b}, energy feedback from star formation uses the thermal implementation of \cite{Vecchia2012} with the heating of a small number of gas particles by a large increment in temperature and the feedback from supermassive black holes (``AGN feedback'') is implemented with a similar method \citep{Booth2009}.



We selected the most massive halo in the simulation at $z=2$. From the three snapshots at redshifts 1, 1.5 and 2, we extracted all SPH particles within a comoving sphere of radius 1.5~Mpc centred on the halo center of potential.
For each SPH particle, we retrieved quantities such as the particle position, velocity, temperature $T$, mass $m$, mass density $\rho$ and chemical abundance\footnote{We made use of the Hydrangea python library: \\ \url{https://hydrangea.readthedocs.io/en/latest/index.html}.}
The main properties of the halo at the selected redshifts are gathered in Table \ref{table:ce22}. The values of the mass within a radius encompassing 500 times the critical density of the Universe at the cluster redshift, $M_{500}$, indicate a well-formed cluster of galaxies at $z=2$ with $M_{500}=7\times 10^{13}$~M$_{\odot}$, equivalent in mass to groups of galaxies in the local Universe. It transitioned into the more massive cluster regime at $z=1$ with $M_{500}=2\times 10^{14}$~M$_{\odot}$. Such system will be the progenitor of a massive cluster of galaxy (e.g., such as the Perseus cluster, at $z=0$). At the redshift of main interest ($z=2$) this cluster is not in a major merging stage. Consistently with other systems at this redshift, this cluster is not relaxed.

\begin{table}[!t]
\caption{Characteristics of Hydrangea halos}
\label{table:ce22}
\centering                          
\begin{tabular}{c c c c c}      
\hline\hline
\\[-0.9em]
$z$  & $M_{500}$ & $R_{500}$ & $T_{500}$ & $\theta_{500}$ \\
& \tiny($10^{14}\,M_{\odot}$) & \tiny(kpc) & \tiny(keV) & \tiny(arcmin) \\
\hline
\\[-0.9em]
1.016 & 2 & 616 & 4.12 & 1.24\\   
1.493 & 1 & 419 & 3.25 & 0.80\\ 
1.993 & 0.7 & 309 & 3.44 & 0.60\\ 
\hline
\end{tabular}
\tablefoot{Characteristics of our target halo from the CE-22 simulation at 3 different redshifts.}
\end{table}


The atomic gas content of the simulation is converted into chemical abundances: for an element $i$, the mass fraction of the element $X_i$, with an atomic mass $\mathcal{M}_i$, is converted into the chemical abundance $Z_i$ in solar metallicity units (as a number density ratio), assuming solar abundance $Z_{\odot,i}$ from \cite{Anders1989}:
\begin{equation}
    Z_i =\frac{X_i}{X_H \times Z_{\odot,i} \times \mathcal{M}_i}
\end{equation}
where $X_H$ is the hydrogen mass fraction of the gas particle.


To serve our show-case study, we investigated various lines-of-sight at each redshift according to the distribution of key physical quantities such as the temperature, density, abundances. 
We qualitatively selected the two projected images presenting (i) the most disturbed and structure-rich (referred to as `irregular'), and, (ii) the most regular and smoothed (referred to as `regular') distributions of these quantities. 
These orientations are used to bracket the impact of projection effects.



\subsection{Model of the ICM X-ray emission}

To model the X-ray emission from our simulated cluster, we down-selected the SPH gas particles to those representative of the ICM in the gas temperature-density plane. We restricted the temperature range to $0.0808 < T_{keV} < 60$. These boundaries are forced by the tabulated X-ray emission model used afterwards (i.e., \texttt{vapec} under \texttt{XSPEC}).
We set an upper limit on the gas density of $n_e < 1\,cm^{-3}$, to avoid overly dense regions in the simulations. Such particles are not representative of the ICM properties, but would nonetheless bias our mock observations. These are likely particles recently affected by the supernovae and/or AGN feedback implementation. We removed about 100 particles out of a few million. 

To model the X-ray emission of the cluster, we followed the procedure described in \cite{Cucchetti2018}.
For each selected gas particle, we assumed the collisionally-ionised diffuse plasma model \texttt{APEC} \citep{Smith2001} computed from the \texttt{AtomDB v3.0.9.} atomic database \citep{Foster2012}. We used its \texttt{vapec} implementation under \texttt{XSPEC} \citep{Arnaud1996} to account for the various chemical abundances traced in the Hydrangea simulations. The aforementioned reference solar abundances are from \citet{Anders1989} and the cross section are taken from \citet{vern96}.

The redshift accounted for  each particle reads as follow:
\begin{equation}
    z = z_{pec} + z_{clus} + z_{clus} \times z_{los} \quad \text{with} \quad 1 + z_{pec} = \sqrt{\frac{1+\beta}{1-\beta}}
\end{equation}

where $z_{pec}$ is linked to the peculiar velocity of the particles along the line-of-sight, $v_{los}$ within the clusters. $\beta=v_{los}/c$, with $c$ being the speed of light and $z_{clus}$ the cosmological redshift of the cluster.

The  X-ray flux is directly proportional to the integral of the electron times proton density, $n_e\times n_H$, over the particle volume, $V$. Following the implementation in \texttt{XSPEC} the normalisation $\mathcal{N}$ of the \texttt{vapec} model writes:
\begin{equation}
    \mathcal{N} =\frac{10^{-14}}{4 \pi [D_A(1+z)]^2}\int_V n_e n_H dV \quad \text{(in cm$^{-5}$)}
\end{equation}
with $D_A$ the angular diameter distance to the source (computed within the cosmological setup of the Hydrangea simulation). We further assume $n_e=1.2\times n_H$. 

We fixed the value of the Galactic hydrogen column density to $0.03 \times 10^{22}$~cm$^{-2}$, a typical value for high galactic latitude. Its associated absorption of soft X-ray photons is modeled with a \texttt{wabs} model \citep{Morrison1983}. 

From these hypotheses, we computed the flux of photons at the Earth emitted by each particle. We used each particle's X-ray spectrum between 0.2-12~keV as a density probability to draw the appropriate number of photons according to a given exposure time and collecting area. Stacked together over all cluster particles, we obtain a photon list at the Earth. The exposure time is systematically fixed to 10 times this of the mock observation exposure in order to provide enough statistics for the instrumental simulations (see Sec.~\ref{sect:mock_observations}.) At this stage the collecting area is chosen to be 20,000~cm$^2$, largely encompassing this of the \emph{Athena} mirrors over the whole energy band. This leads to an oversized list of photons providing proper statistics for the telescope and instrumental simulations and avoid any duplication or under-sampling biases.


\section{Simulated X-IFU/\emph{Athena} observations}
\label{sect:mock_observations}

\subsection{The \emph{Athena} telescope and X-IFU instrument}
\emph{Athena} is the next generation of X-ray telescope from the European Space Agency \citep{bar17}. It implements the science theme of the Hot and Energetic Universe \citep{nan13}. With a focal length of 12\,m it will embark a Wolter-I type mirror initially expected to have a collecting area of 14,000~cm$^2$ at 1~keV, and to provide a spatial resolution of 5~arcsec FWHM. Two instruments will board the payload, the Wide field Imager \citep[WFI --][]{Meidinger2017}, and the X-ray Integral Field Unit \citep[X-IFU --][]{barret2023}.

In this study we focus on the capabilities of the X-IFU instrument, whose main initial high level performance requirements relevant for our study include a spectral resolution of 2.5~eV over the 0.2-7~keV band. This would represent a gain of a factor of $\approx\,$50 with respect to XMM-\emph{Newton}. The effective area (constrained by the mirror collecting area) shall be of 10,000~cm$^2$ at 1~keV, whilst the hexagonal field-of-view shall have an equivalent diameter of 5~arcmin.

We refer the reader to \citet{barret2023} for an comprehensive description of the X-IFU instrument.

\subsection{X-IFU mock observations}
\label{s:obs}

To simulate the observations with the X-IFU instrument, we followed the method described in \citet{Cucchetti2018} and summarised in the following.

We made use of the SImulation of X-ray TElescopes software package \citep[\texttt{SIXTE},][]{Wilms2014,Dauser2019}. SIXTE ingests as input a SIMPUT \citep{Schmid2013} file containing either emission spectra of individual sources or regions, or directly a list of photons at the telescope.  

The SIXTE website\footnote{\url{https://www.sternwarte.uni-erlangen.de/sixte/}} distributes the baseline setup for the X-IFU/\emph{Athena}, as described in the above section and detailed in \citet{barret2023}, formatted for the use of SIXTE. This includes ressources and configurations such as the focal plane geometry, the point spread function (PSF), the vignetting, the instrumental background, crosstalk, the ancillary response file (ARF) and the redistribution matrix file (RMF), as provided by the X-IFU consortium\footnote{\url{http://x-ifu.irap.omp.eu/}}. We used this baseline setup in this work. 

We took into account the astrophysical emission from foreground and background emissions, and co-added them to our cluster of galaxies into a total photon list received at the Earth. We modelled the Galaxy halo and local bubble as diffuse emission according to the model proposed by \citet{McCammon2002}, that is an absorbed (\texttt{wabs*apec}) and unabsorbed (\texttt{apec}) thermal plasma emission model, respectively. 
As per the required \emph{Athena} spatial resolution, 80\% of the sources constituting the extragalactic background are expected to be resolved \citep{Moretti2003}. This corresponds to a lower flux limit for individual simulated sources of $\approx 3\times 10^{-16}$~ergs/s/cm$^2$ for a 100~ksec exposure time  (down scaled to $\approx 10^{-16}$~ergs/s/cm$^2$ for 1~Msec). 
AGN with fluxes below this limit are treated as a diffuse contribution, and modelled with an absorbed power-law spectrum parameterised as in \citet{McCammon2002}. To avoid double-counting resolved AGN, we set the normalisation of this model to 20\% of its nominal value.
The resolved cosmic X-ray background (CXB) is accounted for by adding the contribution of foreground and background AGNs according to the procedure described by \citet{Clerc2018}. In short, AGNs are drawn from their luminosity function, $N(L_{0.5-2keV}, z)$  \citep[from][]{Hasinger2005}. Each source is assigned an emission spectrum from \citet{Gilli2007}. They are then uniformly distributed over the simulated solid angle (thus neglecting any clustering effects).
For both the Galactic and non-resolved CXB emission we used the model parameters provided by \citet{Lotti2014}. 
We fed the total (cluster of galaxies + astrophysical contamination) photon list to the \texttt{xifupipeline} method of \texttt{SIXTE} in order to generate a mock event list. All aforementioned instrumental effects and setups were accounted for, except for the cross-talk, which is irrelevant for our study. 
For each three redshifts and two lines-of-sight orientations we selected for our Hydrangea cluster (see Sec.~\ref{s:hyd} and Table~\ref{table:ce22}), we generated two mock observations of one single pointing with 100 and 250~ksec, respectively. For the cluster at $z=2$, we also ran a third deep exposure of 1~Msec.

\begin{figure}
   \centering
   \includegraphics[width=\hsize]{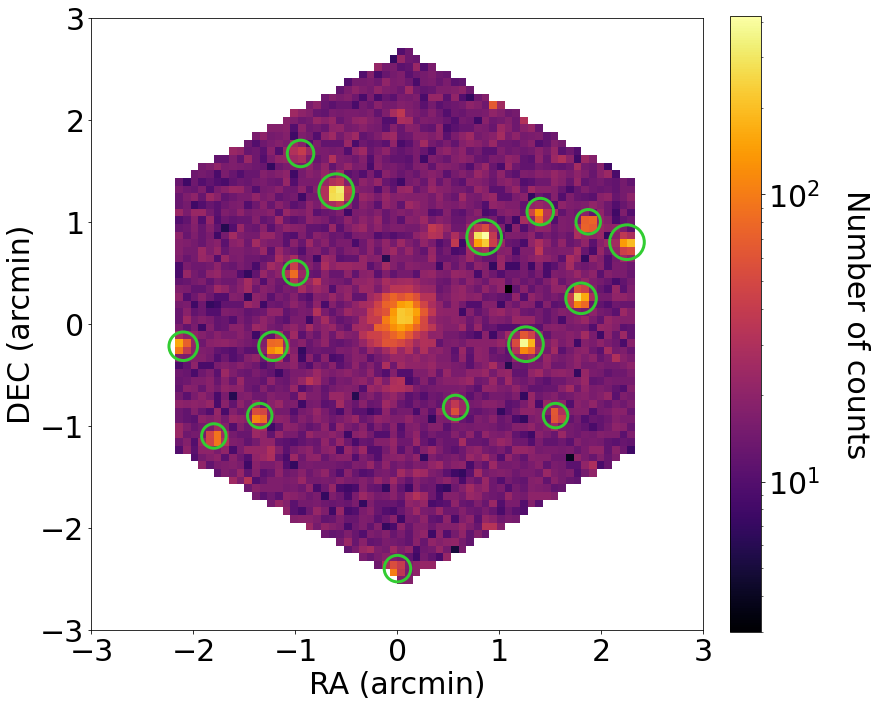}
      \caption{X-IFU map of a 100~ksec raw number count for our Hydrangea cluster of galaxies at $z=2$ and for projection `regular'. The image includes astrophysical and instrumental backgrounds. The green circles mark the loci of the point sources to be excised. 
      }
         \label{Fig:xifucount}
\end{figure}

\subsection{Processing of the mock data}
\label{s:pro}

For the purpose of our study, we assumed in our analysis spherical symmetry for the cluster, and we focus on a radial analysis.

\subsubsection{Point source masking}
We first proceed by generating a raw count map, as shown on Fig.~\ref{Fig:xifucount} for $z=2$ and projection `regular'. As in \citet{Cucchetti2018}, we flagged the loci of the simulated AGNs with fluxes above the limit defined in Sec.~\ref{s:obs}, that is $\approx 3\times 10^{-16}$~ ($\approx 10^{-16}$)~ergs/s/cm$^2$ for an exposure time of 100~ksec (1~Msec) 
They are masked by excluding all the pixels with number counts 2$\sigma$ above the mean count number of the pixels neighbouring their positions. This ensures that the ICM emission is accounted for. This procedure led to an effective flux threshold asymptotically reaching the value of $\approx 3\times 10^{-16}$ ($10^{-16}$)~ergs/s/cm$^2$ with the decreasing ICM emission. We performed a visual check and manually masked any remaining obvious point sources or overdense count regions that could arise from the Hydrangea simulation itself.

\subsubsection{Surface brightness profile}
\label{Sec:sbprofile}

We derived the surface brightness (SXB) profile from the raw count image masked from the point sources in order to account for the geometry of the X-IFU detector. We thus avoid artificially over sampling the intrinsic spatial resolution of the instrument (that is the convolution of the mirror PSF and the pixel size). The focal plane pixels are attributed to the radial annuli which encompass the position of their centre. Hence the pattern of pixels populating the various radial annuli slowly tend to actual geometrical annuli over the detector array with increasing radius. However, the inner annuli {significantly} depart from such a regular shape. For instance, the central annulus contains a single pixel, whilst the second one contains the four nearest adjacent pixels to the central one in a cross pattern. This specific pattern is properly accounted for in the fit of the SXB profile (see Sec.~\ref{s:mod}).

In our mock observation, the position of the cluster projected centre, as defined in the Hydrangea simulation, coincides with the instrument optical axis. In other words, the cluster center is positioned at the centre of the instrument detector array. 
The SXB profile is extracted in the 0.4-1~keV band, where the bremsstrahlung emission of the ICM is still relatively flat as a function of energy even at $z=2$, hence minimising its dependence on the gas temperature. The derived SXB profile for $z=2$ and projection `regular' is presented on Fig.~\ref{fig:result_100ks_z2_grouped_observables}, left panel.


\subsubsection{Spectral analysis}
\label{s:spectral}

To recover the radial distribution of the gas temperature and chemical abundances, we binned our mock event list into six concentric annuli. As for the SXB profile, the centre is chosen to match the position of the projected centre of the simulated cluster.  
The six annuli were defined to gather at least 10,000 counts in order to populate the spectral channels of the X-IFU. The spectra are extracted using the \texttt{makespec} function of \texttt{SIXTE} and binned according to the optimal method by \cite{Kaastra2016}. 
A specific ancillary response is computed per annulus accounting for the mirror vignetting weighted for each contributing pixel by its number counts. 

The local background spectrum is extracted from the area of the field-of-view beyond $R_{500}$. It is fitted with the model described in Sec.~\ref{sect:mock_observations}, that is \texttt{apec+wabs*(apec+powerlaw)}. All parameters but the three normalisations are fixed. 
This best fit background model is then rescaled to the area of each annulus (i.e., the pixel solid angle times the number of pixels belonging to the annulus) and fixed as such for the fit of the ICM model.   
We fitted the spectra of the galaxy cluster with a \texttt{wabs*vapec} model under \texttt{XSPEC} with Cash statistics \citep{Cash1979}, with the gas temperature, the normalisation and the Fe, Si and Mg abundances as free parameters. We limited our investigation of the chemical abundances to these three elements, as they present the most prominent lines with a reasonable probability to be detected out to a redshift of $z=2$. All other abundances were fixed to the average value of the projected emission-measure weighted abundances from the Hydrangea halo particles over a given annulus.


The reconstructed temperature and abundances radial distribution are shown in Fig.~\ref{fig:result_100ks_z2_grouped_observables}, right panel and Fig.~\ref{fig:result_100ks_z2_grouped_abundances}, respectively.


\section{Forward-modelling and MCMC analysis}
\label{s:mod}


We chose a forward modelling approach to fit the SXB, temperature and abundance profiles in order to reconstruct the 3D radial distribution of the thermodynamical and chemical properties of the simulated cluster of galaxies.
We recall that we assume spherical symmetry for our simulated clusters. 

\subsection{The forward-modelling procedure}
\label{s:for}


We considered the gas density and pressure as two independent physical quantities in our model.

We modelled the density distribution with a `simplified' Vikhlinin functional \citep{Vikhlinin2006}:

\begin{equation}
\label{eq:density_profile}
    n_e^2(x) = \frac{n_0^2}{(x/r_s)^{\alpha_1}[1+(x/r_s)^2]^{3\beta_1 - \alpha_1/2}}
\end{equation}
where $x=r/R_{500}$. $n_0$ is the central density, $\alpha_1$ and $\beta_1$ the inner and outer slopes, respectively, and $r_s$ the scale radius.

The pressure radial distribution is described by a gNFW profile \citep{Nagai2007}
\begin{equation}
\label{eq:pressure_profile}
    P(x) = \frac{P_0}{(c_{500}~x)^{\gamma}[1+(c_{500}~x)^{\alpha_2}]^{(\beta_2-\gamma)/\alpha_2}}
\end{equation}
Where $P_0$ is the central pressure, $c_{500}$ is the concentration at a radius of $R_{500}$, $\alpha_2$ and $\beta_2$ the intermediate and outer slopes, respectively. We do not have enough leverage with our observational constraints to fit $\gamma$, the inner slope. We thus let it as a free parameter with a Gaussian prior with mean $0.43$ and standard deviation of $0.1$, the values derived from the XCOP sample \citep{Ghirardini2019}.

The temperature distribution is simply derived from the gas density and pressure assuming the ICM to be a perfect gas with  $P = n_e\times k_B T$ ($k_B$ being the Boltzmann constant). The entropy is reconstructed assuming the conventionally adopted relation in X-ray astronomy as introduced by \citet{Voit2005}, that is $K = k_B T/{n_e^{2/3}}$.

For the radial distributions of chemical elements, we adopted a simple power-law model:
\begin{equation}
\label{eq:abundance_profile}
    A(x) = A_{0}~x^{-p}
\end{equation}
where $A_{0}$ is specific to each element. However we assumed the same slope $p$, for all elements, considering it a fair assumption according to \cite{Mernier2017}.

The remaining twelve free parameters are listed in Table~\ref{table:prior} together with their initial values and priors adopted in our MCMC fit.
~\\

\begin{figure}
    \centering
    \includegraphics[width=\linewidth]{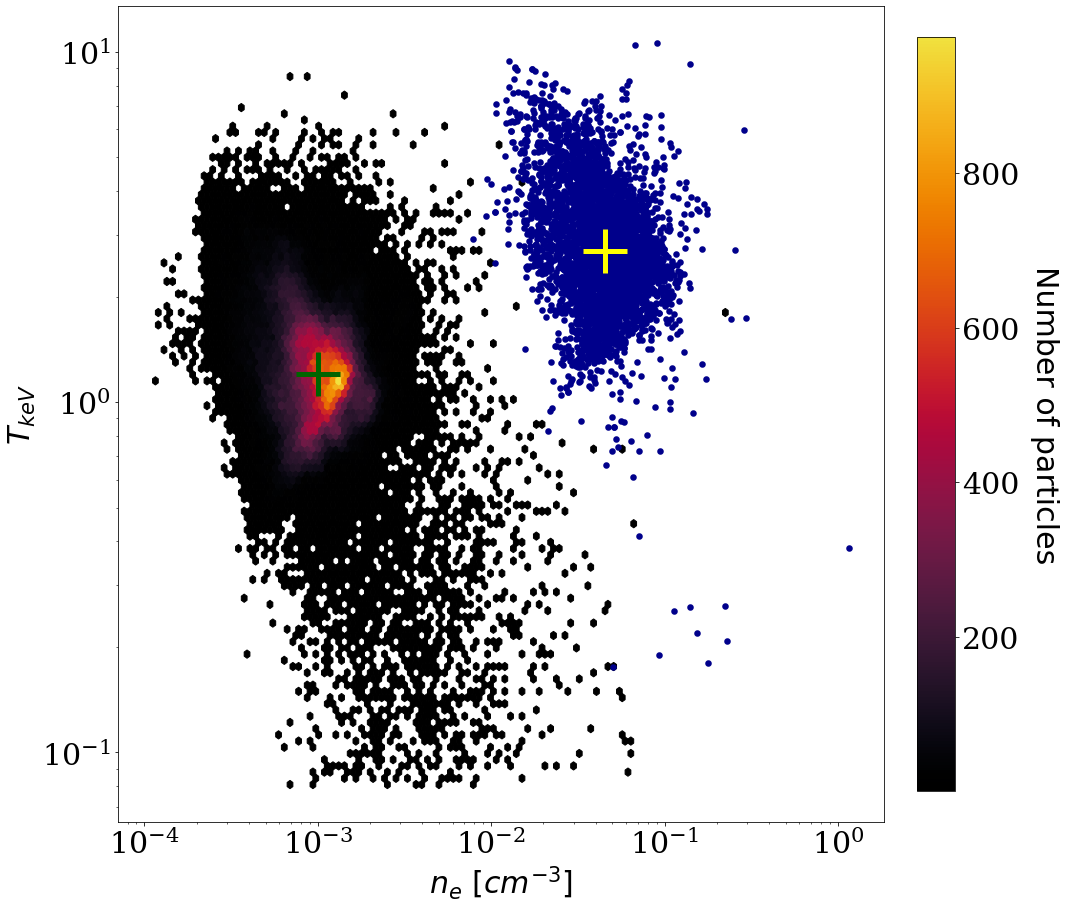}
    \caption{Dispersion of thermodynamic quantities (electronic density $n_e$, units cm$^{-3}$ and temperature $T$, units keV) in one radial shell with $300$\,kpc $<r<$ 310\,kpc of the $z=2$ cluster. The leftmost cloud is a density map of extracted particles from the Hydrangea-simulated cluster. The leftmost green cross indicates the position of the median density and temperature.
    The cloud of 5000 blue points on the right-hand side illustrates our random generation of a new model, when requesting a new set of median values (as shown by the yellow rightmost cross). In this process, the shape of the dispersion in the $\log-\log$ plane is essentially maintained fixed and translated.}
    \label{fig:dispersion_shifting}
\end{figure}

We account for the parameter dispersion in each spherical shell by randomly drawing 5000 particles in the shell and adopting their values in density and pressure around the  shell fiducial value. We aim to compare our results with volume-weighted distributions of the thermodynamic quantities (Sect.~\ref{sect:results}), so as to minimise the impact of high-density, small-volume particles~; therefore we weigh the random draws by the volume of each particle. In doing so, we follow the dispersion of thermodynamic quantities predicted by the Hydrangea simulation about their fiducial profile. Fig.~\ref{fig:dispersion_shifting} illustrates this process and shows how translating the cloud of particles in the $\log n_e-\log T$ plane provides a new distribution of densities and temperatures dispersed around a newly requested median value. These distributions are clipped at zero and at the maximal values allowed in our setup (1\,cm$^{-3}$ and 15\,keV). This process leads to 5000 spherically symmetric models, each contributing in equal proportion to the final model. Taken individually, they do not serve as realistic descriptions of the cluster~; they serve as intermediate models, enabling propagation of the scatter in thermodynamic profiles.

We compute the X-ray emission of the `toy model' cluster by averaging that of the 5000 models together.
We apply an Abel transform, using the PyAbel\footnote{\url{https://pyabel.readthedocs.io/en/latest/}} python package with the Hansenlaw transform method \citep{Hansen1985}, to each of the 5000 models and average the resulting surface brightness profiles. Assuming circular symmetry, the profile is distributed into a two-dimensional grid generously oversampling the X-IFU pixel size. Emission-measure-weighted temperature and abundance profiles are constructed similarly. 
We finally convolve these parameter grids with the PSF kernel and reshape it to the pixelization and geometry of the X-IFU focal plane. We reproduce the source and pixel masking applied to the mock data. With this process we ensure faithfully reproducing the input image characteristics (see Sec.~\ref{s:pro}).

\begin{table}[!t]
\caption{Cluster model parameters}
\label{table:prior}      
\centering                          
\begin{tabular}{l l l l}      
\hline\hline
\\[-0.9em]
Parameter & Initial value & Priors & Unit \\    
\hline
\\[-0.9em]
   $n_0$ & $10^{-2}$ & $\mathcal{U} (10^{-7} ~-~ 1)$ & cm$^{-3}$\\   
   $\alpha_{1}$ & $-0.3$ & $\mathcal{U} (-1 ~-~ 3)$ & \\ 
   $\beta_{1}$ & $0.9$ & $\mathcal{U} (0.1 ~-~ 4)$ & \\ 
   $r_{s}$ & $0.3$ & $\mathcal{U} (0.1 ~-~ 1)$ & $R/R_{500}$ \\ 
   $P_0$ & $5 \,10^{-2}$ & $\mathcal{U} (0 ~-~ 0.2)$ & keV\,cm$^{-3}$\\ 
   $\alpha_{2}$ & $3$ & $\mathcal{U} (0 ~-~ 4)$ & \\ 
   $\beta_{2}$ & $5.17$ & $\mathcal{U} (1 ~-~ 10)$ & \\
   $\gamma$ & $0.43$ & $\mathcal{N} (0.43,0.1)$ & \\
   $c_{500}$ & $2.4$ & $\mathcal{U} (1 ~-~ 4)$ & \\
   $A_{0,Fe}$ & $5 \,10^{-2}$ & $\mathcal{U} (0 ~-~ 2)$ & $Z/Z_{\odot}$ \\
   $A_{0,Si}$ & $0.13$ & $\mathcal{U} (0 ~-~ 2)$ & $Z/Z_{\odot}$ \\
   $A_{0,Mg}$ & $9 \,10^{-2}$ & $\mathcal{U} (0 ~-~ 2)$ & $Z/Z_{\odot}$ \\
   $p$ & $0.6$ & $\mathcal{U} (-1 ~-~ 3)$ & \\[0.2em]
   \hline
   \\[-0.9em]
   $bkg$  & $10^{-2}$  & $\mathcal{U} (0 ~-~ 0.3)$  & cts/s/arcmin$^2$\\
\hline                                  
\end{tabular}
\tablefoot{Initial values and priors of the 13 free parameters describing our cluster model in the MCMC fit. The last, fourteenth, parameter is also let free and accounts for the level of uniform background in the 0.4 -- 1\,keV image.}
\end{table}

\subsection{The MCMC fit}
We simultaneously fitted our SXB, temperature and abundances profiles using a Bayesian MCMC approach. The modeling procedure described above (including the assessment and propagation of the parameter dispersion in shell) is reproduced at each step of the MCMC. 

We formulated the associated likelihood as follows:
\begin{equation}
        \chi^2 = -2~log \mathcal{L} = \sum_{p_j} \sum_{i} \frac{(y_{i,p_j} - M_{i,p_j})^2}{\sigma_{i,p_j}^2}
\label{eq:logl}
\end{equation}
where $y_{i,p_J}$, $\sigma_{i,p_j}$ and $M_i$ are the mock data, associated mock error and model for profile $p_j$, respectively. $p_j$ denotes the set of the SXB, temperature and Mn, Si, Fe abundances profiles. Such a likelihood implicitly assumes normal-distributed uncertainties. In case of asymmetric uncertainties in the observable, we take $\sigma_{i, p_j}$ as the arithmetic mean of the upper and lower bounds of the 68\% confidence level.

Our MCMC sampling of the parameter space made use of python-based code \texttt{emcee} \citep{Foreman-Mackey2019}.

\subsection{Validation} 
\label{sect:validation}

Before proceeding to the analysis of mock Hydrangea data, we first validate our fitting procedure using a simpler, spherically symmetric model. Its physical quantities (density, pressure, element abundances) are drawn following parametric profiles (Eq.~\ref{eq:density_profile}, \ref{eq:pressure_profile} and~\ref{eq:abundance_profile}). The numerical values are chosen so as to approximately reproduce the radial behaviour of the $z=2$ cluster that we picked in the Hydrangea simulation (Table~\ref{table:ce22}). In contrast with the Hydrangea cluster though, we impose that each of these quantities, taking a single value within a spherical shell at radius $r$. This idealised model is transformed into a mock 100\,ks X-IFU observation (as described in Sect.~\ref{sect:mock_observations})~; this step includes the extraction of temperature and surface brightness profiles.
We then fit the X-IFU mock observations with the exact same cluster model used to fit the actual simulation, letting 14 parameters freely evolve~; namely: $n_0$, $\alpha_1$, $\beta_1$, $r_s$, $P_0$, $\alpha_2$, $\beta_2$, $\gamma$, $c_{500}$, $A_{0, i}$ ($i$ being one of Fe, Si and Mg), $p$ and finally a uniform background level in the 0.4--1\,keV imaging band. Priors and initial values for the 14 parameters are listed in Table~\ref{table:prior}.

Figure~\ref{fig:validation} shows the results of our validation test. We find overall good agreement between the input model and the inferred properties. The three-dimensional density and pressure profiles roughly agree with the input profiles within the 68\% confidence intervals, at least across the radial range of applicability of our procedure (that is, at radial distances located between the PSF size, $r \approx 0.1 R_{500}$, and the outermost radius, $R_{500}$). Nevertheless, significant deviations appear, especially when considering the two-dimensional temperature profiles and the comparison between the purple and black lines in the top-right panel of Fig.~\ref{fig:validation}. Despite the simplicity of the input model, projection effects along the line of sight induce temperature and abundance mixing, which limit the ability of a single APEC model to account for the observed spectrum. This effect leads to underestimated uncertainties issued by the \texttt{XSPEC} fit (green error bars). In addition to this issue comes our (so far unverified) assumption that weighting the 3-d temperature by the emission-measure provides a fair representation of the 2-dimensional temperature profile.
We have checked that such deviations are not solely of statistical origin by increasing the exposure time of the mock observation to 1\,Ms and finding similar offsets in the projected temperature profile (Fig.~\ref{fig:validation_1Ms}). However, repeating the experiment with a toy-model cluster placed at $z=1$ provides better agreement in the recovered profiles (Fig.~\ref{fig:validation_z1}). Indeed, the larger apparent $R_{500}$ as compared to the $z=2$ system enables defining a finer two-dimensional binning of the temperature profiles (ultimately limited by the X-IFU pixel and/or PSF size), hence mitigating the line-of-sight mixing effects.

In summary, our validation tests demonstrate the ability of our procedure to recover input profiles. However, we highlighted a limitation due to line-of-sight mixing of temperature (and abundances), attributed to both the finite X-IFU angular resolution relative to the $z=2$ cluster angular span, and to the slight inadequacy of the spectral fitting model. These deviations propagate into the inference of the 3-dimensional thermodynamic profiles and limit our ability to recover their exact distributions. These limitations are in fact inherent to any observation of multi-phase diffuse gas, irrespective of the instrument in use.

\section{Results}
\label{sect:results}

We now present the results obtained by fitting the $z=2$ galaxy cluster extracted from the Hydrangea simulations (Sect.~\ref{sect:simulations}) and folded through the X-IFU instrumental response assuming a 100\,ks exposure time. Similarly as for our validation model (see Sect.~\ref{sect:validation}), we fit mock data with a spherically symmetric model with 14 free parameters. They are listed in Table~\ref{table:prior}. They enter Eqs.~\ref{eq:density_profile} to~\ref{eq:abundance_profile} and govern the median 3-dimensional model for $n_e(r)$, $P(r)$ and $A(r)$. The thirteenth parameter is an additional background level in the $0.4-1$\,keV imaging band. Accounting for the intrinsic dispersion of these quantities around their median values requires prior knowledge of its radial behaviour (or a parameterised model thereof). We simply assumed that the dispersion in the $(\log n_e, \log T)$ plane follows that of the Hydrangea cluster in each radial shell (see Fig.~\ref{fig:dispersion_shifting} and Sec.~\ref{s:for}). This assumption avoids introducing additional model parameters, although one expects that using the exact dispersion for the forward-model may put our results on the optimistic side. For computational reasons, we do not assume any dispersion of the abundance within a spherical shell in our forward model.

\begin{figure*}
\begin{center}
    \begin{tabular}{cc}
    \includegraphics[width=0.48\linewidth]{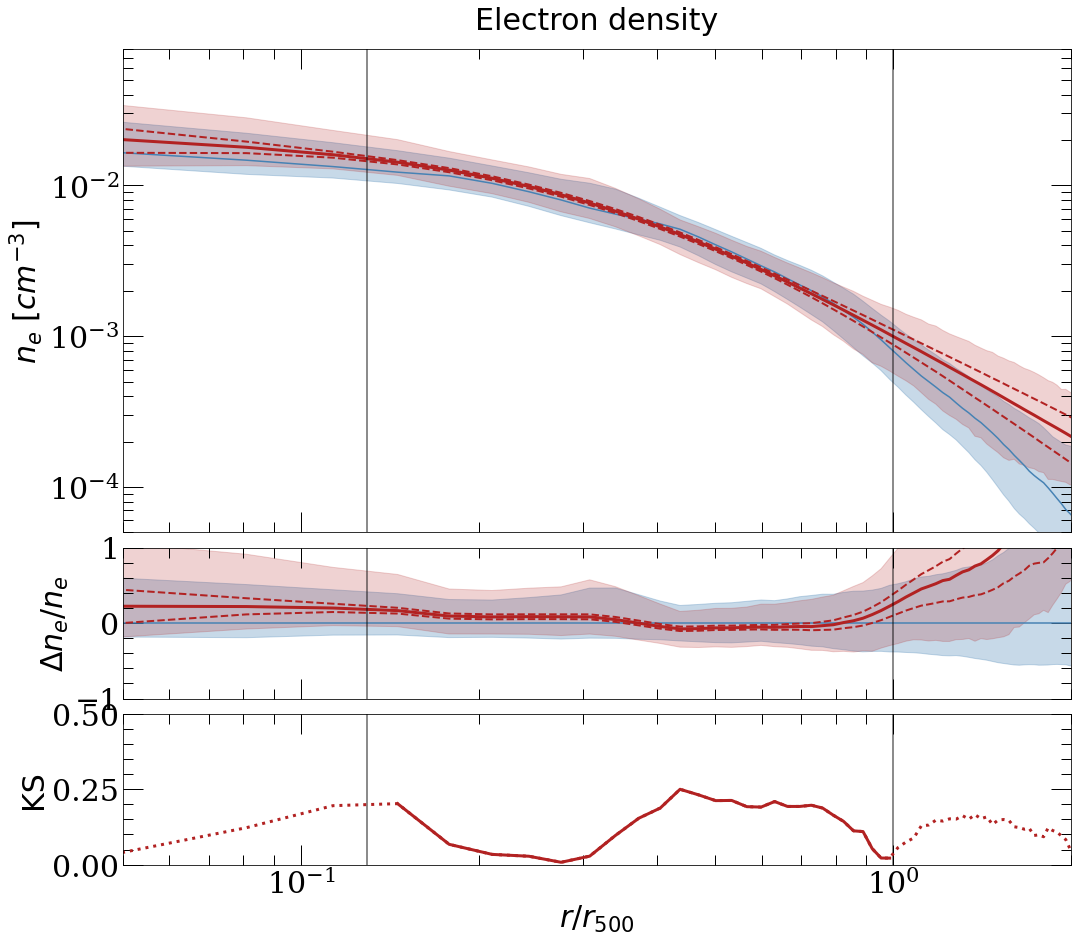} &
    \includegraphics[width=0.48\linewidth]{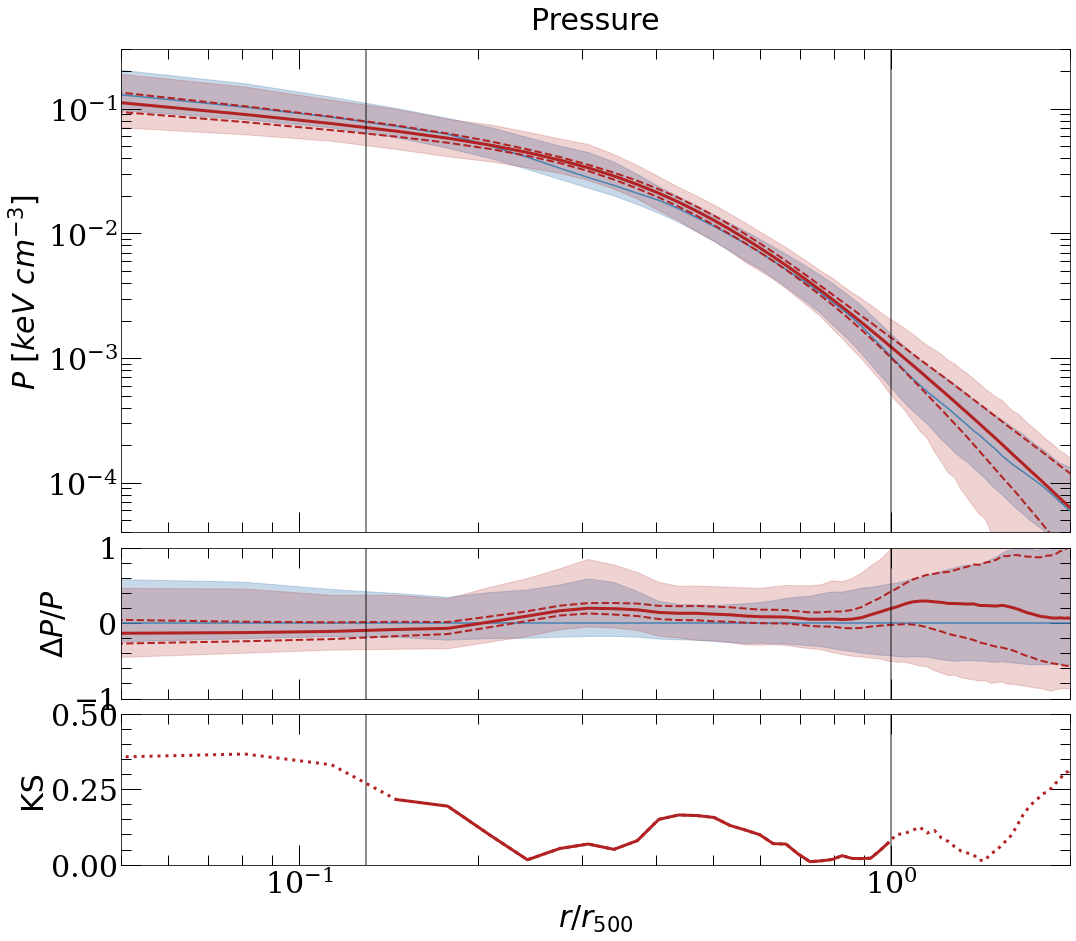} \\
    \includegraphics[width=0.48\linewidth]{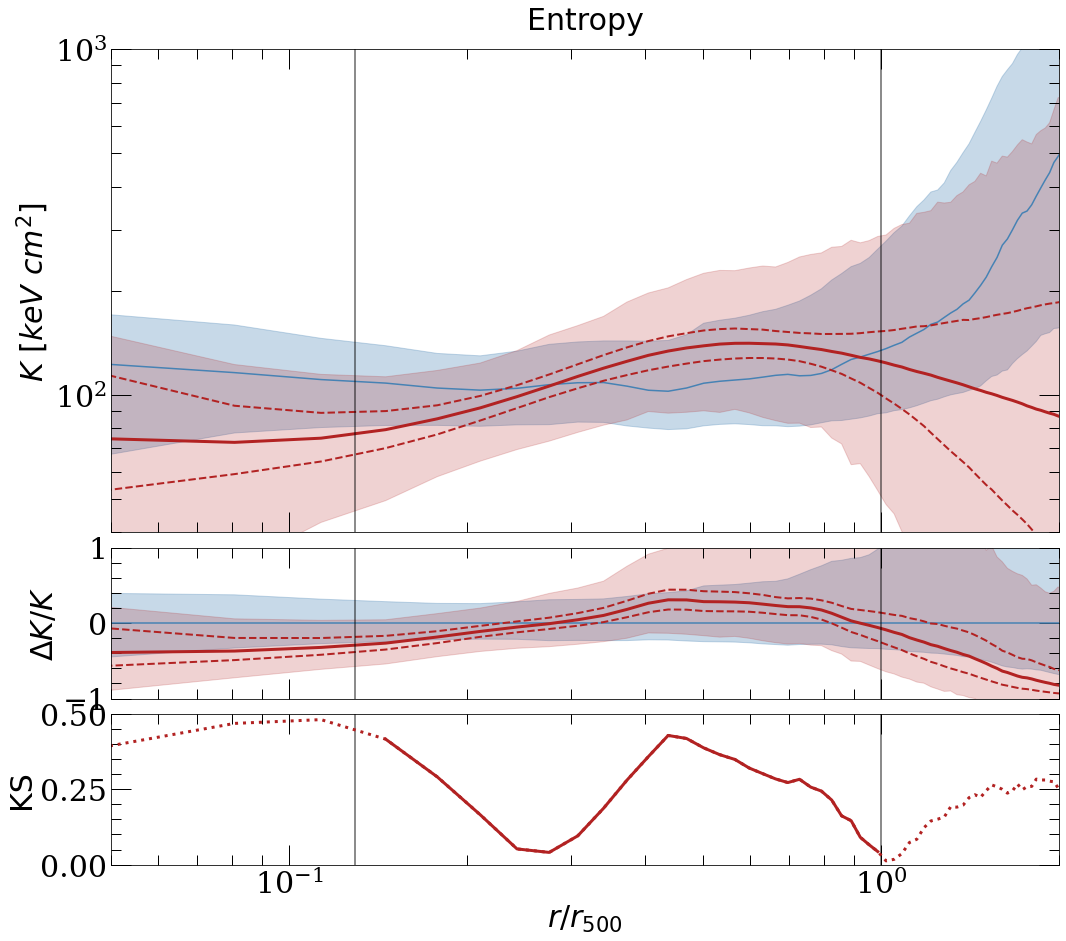} &
    \includegraphics[width=0.48\linewidth]{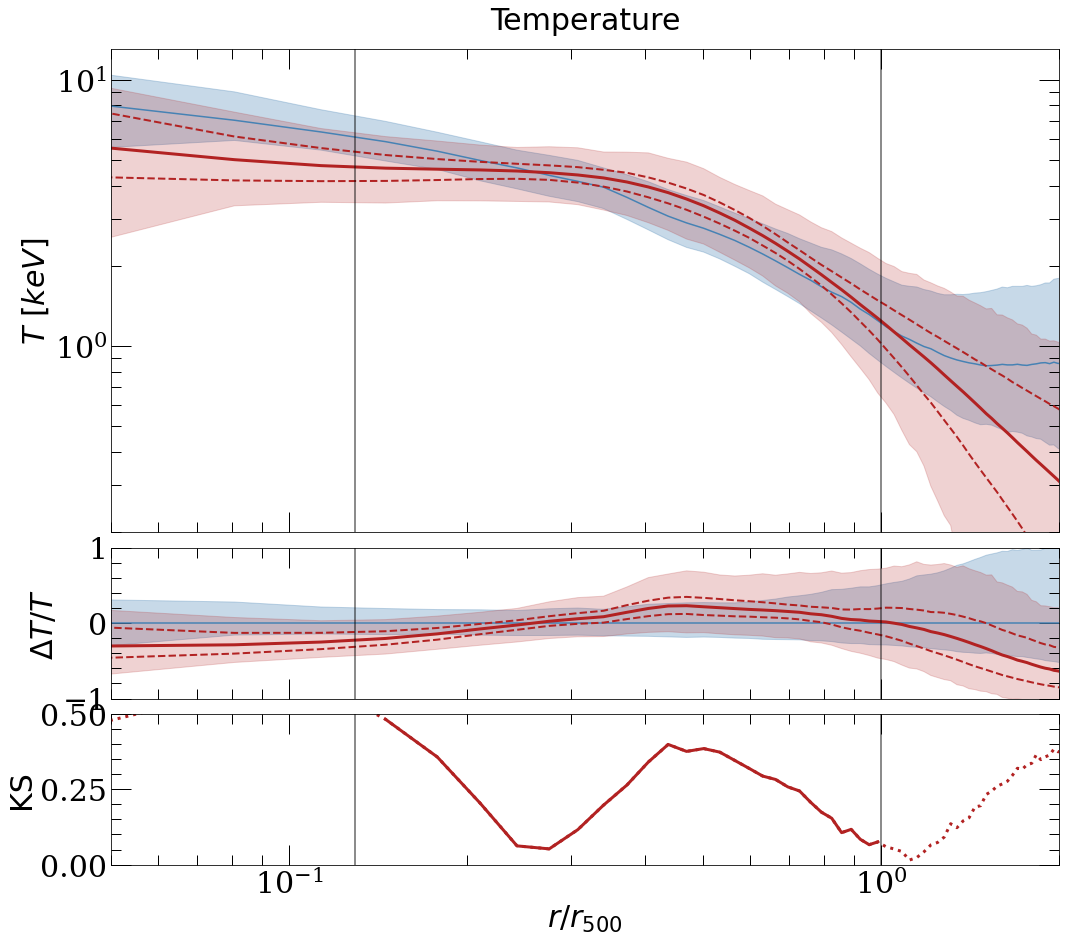} \\
    \end{tabular}
\caption{Three-dimensional thermodynamic quantities (in blue) for the $z=2$ galaxy cluster with projection `regular' in the Hydrangea sample, and their best-fit models inferred from an X-IFU 100\,ks exposure (in red). Each panel representing one quantity is made of three plots. The top curves display the radial profiles~; the blue shaded envelope represents the dispersion of the gas particles in the hydro-simulation. The red dashed lines indicates the effect of the variance of the 14 free model parameters~; the shaded envelope also includes the radial dispersion encapsulated in our model.
The middle plot represents the deviation of the profile relative to the input median profile. The bottom plot shows the results of a Kolmogorov-Smirnov (KS) test performed at each radius $r/r_{500}$, related to the probability that the input and best-fit profiles do not originate from the same distribution (the lower KS, the closer the agreement between the profiles).
Vertical lines indicate the range of applicability of our modelling procedure.
\label{fig:result_100ks_z2_grouped_physical}
}
\end{center}
\end{figure*}

\begin{figure*}
\begin{center}
    \begin{tabular}{cc}
    \includegraphics[width=0.465\linewidth]{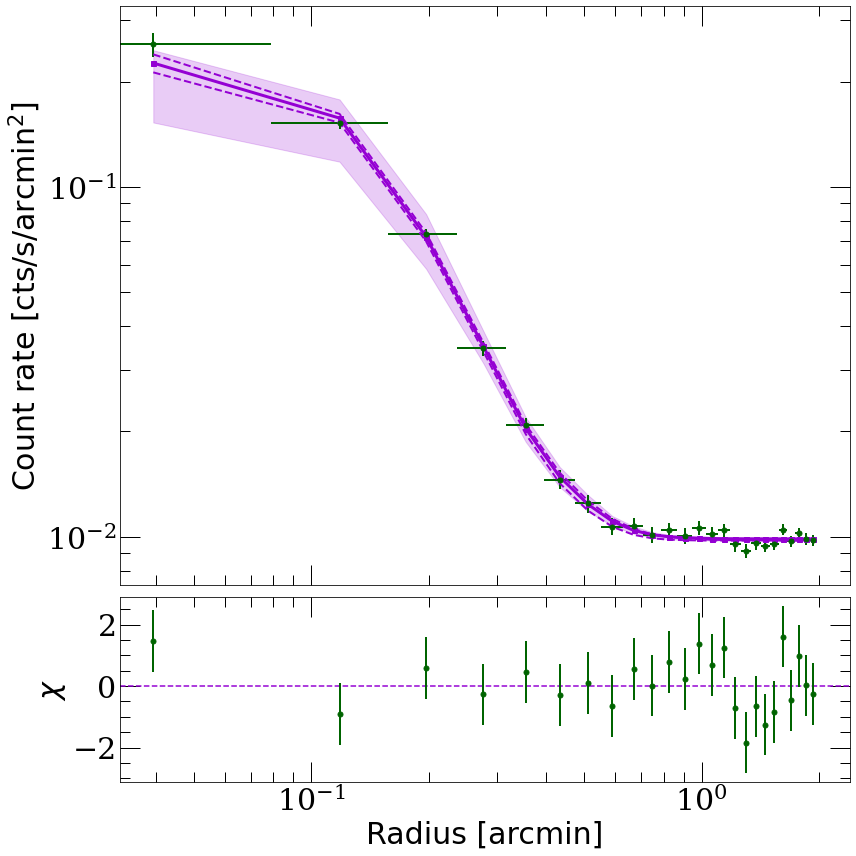} &
    \includegraphics[width=0.48\linewidth]{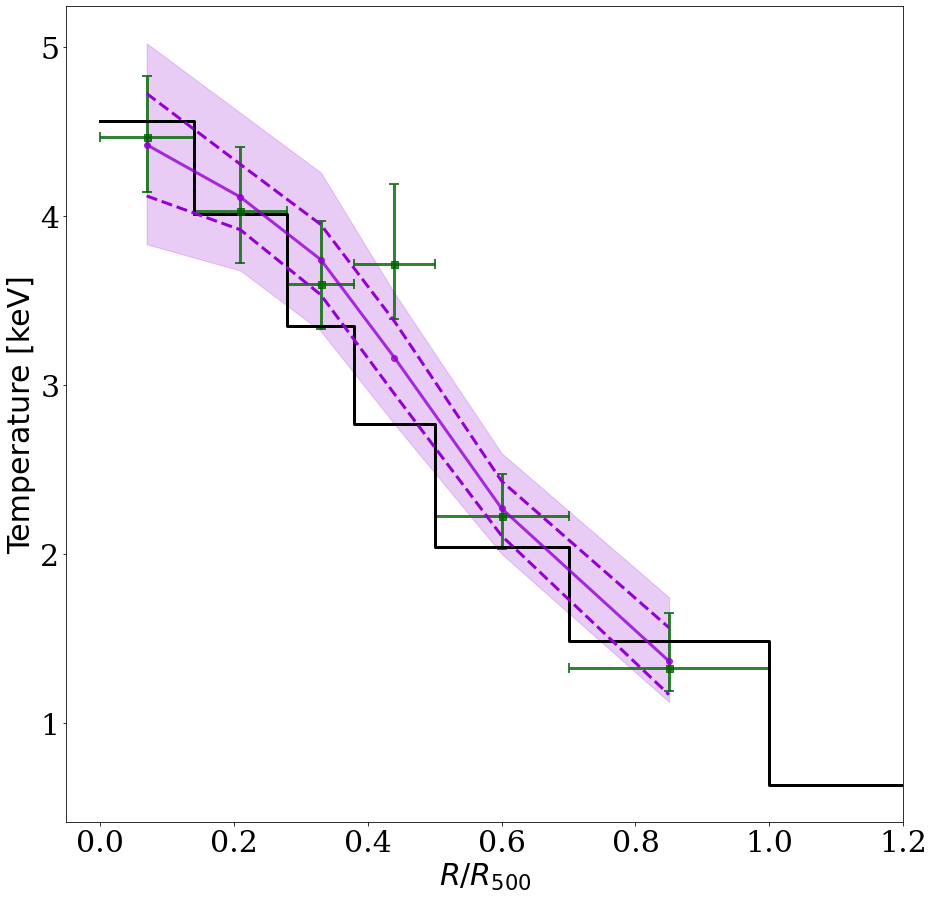} \\
    \end{tabular}
\caption{Projected observables and their best-fit models for the $z=2$ galaxy cluster with projection `regular' in the Hydrangea sample, as seen by X-IFU in a 100\,ks exposure.
Left: surface brightness profile in the 0.4--1\,keV band and its uncertainties (green points). The posterior mean (`best-fit model') and its 68\% confidence envelope are represented in purple colours. The bottom panel displays the deviation of the measurements relative to the best-fit, normalised by the errors.
Right: two-dimensional temperature profile as measured by \texttt{XSPEC} in each annular bin (green points and errors). The emission-measure weighted temperature $T_{EM}$ (known from the hydrodynamic simulation) is shown as a thick black line.
In both panels, the shaded purple envelope includes both the contribution of the intrinsic dispersion of physical quantities within the radial shells and the propagated variance of the 14 free model parameters (dashed lines).
\label{fig:result_100ks_z2_grouped_observables}
}
\end{center}
\end{figure*}

\begin{figure*}
\begin{center}
\includegraphics[width=0.95\linewidth]{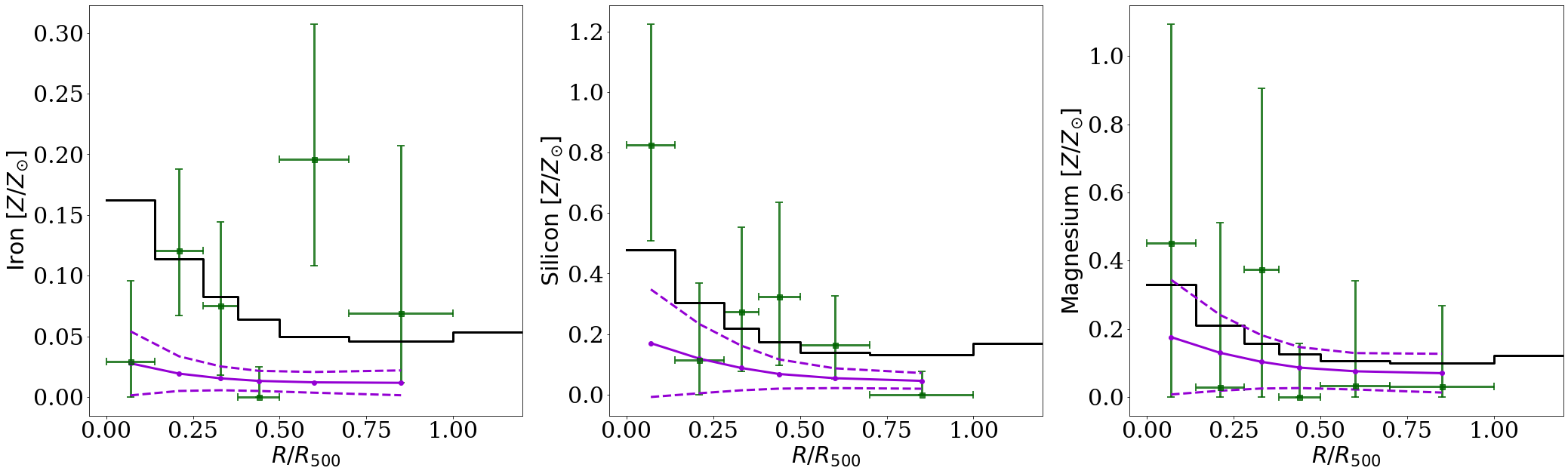}
\caption{Projected abundance profiles and their best-fit models for the $z=2$ galaxy cluster with projection `regular' in the Hydrangea sample, as seen by X-IFU in a 100\,ks exposure. Each panel corresponds to one chemical element of Fe, Si and Mg. The measurement output by \texttt{XSPEC} is shown as green points (and errors). The emission-measure weighted abundance profile (known as input) is displayed with a thick black line. The best-fit model is displayed in purple and the dashed lines represent the error of the free model parameters.
\label{fig:result_100ks_z2_grouped_abundances}
}
\end{center}
\end{figure*}

Figures~\ref{fig:result_100ks_z2_grouped_physical},  \ref{fig:result_100ks_z2_grouped_observables} and~\ref{fig:result_100ks_z2_grouped_abundances} illustrate the outcome of the forward-modelling procedure. In each figure the uncertainty on the 14 free model parameters is propagated and provides the envelope indicated with red dashed lines. In contrast to the validation model (e.g.~Fig.~\ref{fig:validation}), our visualization now incorporates the effect of the intrinsic dispersion in the thermodynamic quantities, which is a key ingredient in our model. In each figure (but for the chemical abundances) this intrinsic dispersion is added in quadrature to the parameter uncertainties in order to provide the red shaded envelope. 
Profiles built from simulation particles also incorporate intrinsic dispersion. In each radial shell we computed the particle-volume weighted histogram of a given quantity (e.g.,~density). Fig.~\ref{fig:result_100ks_z2_grouped_physical} reports the associated 16, 50 and 84 percentiles as a plain blue line and a blue shaded envelope.

Focusing first on the recovery of physical thermodynamic quantities (electron density $n_e$, pressure $P$, temperature $T$ and entropy $K$ in Fig.~\ref{fig:result_100ks_z2_grouped_physical}) we notice a qualitatively good agreement between the profiles recovered from the fit and the input profiles. The large amount of intrinsic dispersion within each spherical shell makes comparison of the median profiles cumbersome, since there is no one unique density (resp. pressure, temperature, entropy) at a given radius, rather a distribution of densities (resp. $P$, $T$, $K$). In order to quantify the agreement between the input and fitted profiles, we performed a two-sample Kolmogorov-Smirnov (KS) test in thin shells at each radius $r$, by comparing the distribution of density values $n_e$ (resp. $P$, $T$, $K$) of the Hydrangea simulation to the distribution of densities (resp. $P$, $T$, $K$) sampled by the MCMC. The KS-statistic is related to the probability to reject the null hypothesis, being that both distributions originate from the same (unknown) distribution. The lower the KS value, the more confident we are that both profiles are in agreement with each other. We also computed the average value of the KS statistics over the radial range comprised between the PSF size and $R_{500}$ and we listed the values in Table~\ref{table:KS} (in bold characters).

Folding the MCMC parameter samples through our model provides the distribution of projected observables, namely the 0.4--1\,keV surface brightness (Fig.~\ref{fig:result_100ks_z2_grouped_observables}, left) and the emission-measure-weighted temperature (Fig.~\ref{fig:result_100ks_z2_grouped_observables}, right).
Most of the dispersion in surface brightness posterior samples originates from the intrinsic dispersion of thermodynamic quantities (mostly $n_e$), while statistical uncertainties on the profiles arising from the MCMC have little impact on the global error budget. The deviation of the reconstructed surface brightness profile, relative to the statistical error, is at most 1.4.
The variance in the posterior projected temperature is roughly equally shared between intrinsic dispersion and parameter uncertainties. Similarly as for the validation case (Sect.~\ref{sect:validation} and Fig.~\ref{fig:validation}), some deviations appear between the best-fit and input models, although they are contained within the 1-$\sigma$ envelope. A noticeable outlier is the \texttt{XSPEC}-fitted temperature in the fourth radial bin at $R \simeq 0.45\,R_{500}$. Part of this bias is due to mixing effects along the line-of-sight, and a spectral model that is unable to account for such mixed components. The bias is also caused by a relatively faint CXB point source located within the brighter cluster region. It is therefore absent from the set of excised points sources (circles in Fig.~\ref{Fig:xifucount}). This unmasked point source brings an excess of high energy photon in the fourth radial bin spectrum, that is sufficient to bias high the \texttt{XSPEC} measurement. The deviation of the reconstructed temperature, relative to the statistical error, is at most 2.9 and at most 1 if we remove the failed measurement in the fourth annulus.

The inference of chemical abundance profiles is depicted in Fig.~\ref{fig:result_100ks_z2_grouped_abundances}. None of the profiles is correctly recovered, in other words, the 68\% posterior confidence level (dashed purple envelope) does not reproduce well the EM-weighted abundance profile (black thick line) known as input from the hydrodynamical simulation. It is worth noting how \texttt{XSPEC}-fitted abundances scatter widely around the expected values, denoting both a lack of statistics and inadequacy in the spectral model due to the mixing of components. The apparent underestimation of posterior abundance profiles originates from the low signal-to-noise ratios and to an improper use of a Gaussian likelihood (Eq.~\ref{eq:logl}) for strictly positive abundance values.
We have verified that increasing the X-IFU exposure to 1\,Ms provides a decent recovery of the iron (Fe) and silicon (Si) profiles, while magnesium (Mg) still suffers from poor statistics (see Sect.~\ref{sect:discussion} and Fig.~\ref{fig:result_1Ms_z2_grouped_abundances}).
Such a result is not surprising due to the faintness of the cluster emission, and to the large dispersion of abundances along a single line of sight in the Hydrangea simulation (that is spatially correlated with the density and temperature) and thus the inability of our model to adequately capture the 3-dimensional structure of the element abundances in such a complex object. This issue is exacerbated by our choice of a crude spatial binning, mostly targeted towards temperature extraction \citep[see e.g.,][presenting an alternative binning scheme tailored for abundance measurements]{Cucchetti2018}.

\begin{table}
\caption{Kolmogorov-Smirnov test results}
\label{table:KS}
\centering                          
\begin{tabular}{l l l c c c c}      
\hline\hline
\\[-0.9em]
 & & [ksec] & $n_e$ & $T$ & $P$ & $K$ \\
 \hline
\\[-0.9em]
\multirow{4}{3em}{$z=1$} & \multirow{2}{3em}{`regular'} & 100 & 0.152 & 0.157 & 0.121 & 0.168 \\
 & & 250 & 0.107 & 0.158 & 0.117 & 0.127 \\
 & \multirow{2}{3em}{`irregular'} & 100 & 0.072 & 0.084 & 0.102 & 0.065 \\
 & & 250 & 0.079 & 0.152 & 0.150 & 0.122 \\
\hline
\\[-0.9em]
\multirow{4}{3em}{$z=1.5$} & \multirow{2}{3em}{`regular'} & 100 & 0.119 & 0.167 & 0.142 & 0.188 \\
 & & 250 & 0.105 & 0.116 & 0.100 & 0.159 \\
 & \multirow{2}{3em}{`irregular'} & 100 & 0.088 & 0.243 & 0.149 & 0.235 \\
 & & 250 & 0.060 & 0.113 & 0.088 & 0.112 \\
\hline
\\[-0.9em]
\multirow{6}{3em}{$z=2$} & \multirow{3}{3em}{`regular'} & {\bf 100} & {\bf 0.136} & {\bf 0.234} & {\bf 0.080} & {\bf 0.241} \\
 & & 250 & 0.431 & 0.368 & 0.614 & 0.387 \\
 & & 1000 & 0.161 & 0.332 & 0.140 & 0.321 \\
 & \multirow{3}{3em}{`irregular'} & 100 & 0.113 & 0.327 & 0.220 & 0.291 \\
 & & 250 & 0.097 & 0.309 & 0.305 & 0.217 \\
 & & 1000 & 0.048 & 0.269 & 0.229 & 0.207 \\
\hline
\end{tabular}
\tablefoot{Radial average of Kolmogorov-Smirnov test statistics, for each of the four thermodynamic profiles recovered by our fitting procedure. The KS tests are computed with respect to the input profiles from the Hydrangea simulation (lower KS values indicate closer agreement). The radial average is evaluated between the size of the PSF and $R_{500}$. Bold characters refer to the configuration specifically discussed in Sect.~\ref{sect:results} and shown in Fig.~\ref{fig:result_100ks_z2_grouped_physical}.}
\end{table}

\section{Discussion}
\label{sect:discussion}

\subsection{On inferring the properties of a $z=2$ cluster of galaxies}

Our analysis demonstrates the capability to infer the (volume-weighted) thermodynamic properties of a realistic cluster of galaxies located at $z=2$, using a moderate exposure budget of 100\,ks with X-IFU onboard \emph{Athena}. Despite the compact faint appearance of this low-mass object at such large cosmological distances (Fig.~\ref{Fig:xifucount}) the effective radial range accessible to X-IFU spans almost one decade between $\approx 0.1-1$\,$R_{500}$. This enables recovery of the shape, amplitude and characteristic slopes of the profiles. The finite instrumental angular resolution prevents accessing smaller scales and resolving core properties.

The density and pressure profiles are the quantities best reconstructed in comparison to the input data, with deviations of the median profile reaching at maximum 20\%. Temperature and entropy display more noticeable deviations, up to 50\% when considering the median profiles. The explanation for this difference relates to the fact that temperature and entropy are deduced from the other two. Therefore, their uncertainties propagate the errors of both density and pressure profiles. Moreover, this difference is also related to our primary observables used for inference. Since surface brightness profiles are measured with much less uncertainty than projected temperature (see e.g.~Fig.~\ref{fig:result_100ks_z2_grouped_observables}), quantities heavily dependent on temperature (temperature itself, and entropy) are much more strongly affected by measurement systematics than density and pressure.

Our $z=2$ study highlights a key point, being that the ICM is not spherically symmetric, nor is it homogeneous within a given radial shell, making a mere comparison of median profiles incapable of grasping the complete reality of the scientific problem. For this reason we have also compared distributions of thermodynamic quantities at fixed radius. Our radially-dependent KS-statistic test indicates again that the recovered density and pressure relate well to the Hydrangea simulation, with KS statistics spanning values between $0$ and $0.2$. Entropy and pressure may display KS-values up to $0.4-0.5$, notably in the centre where PSF blurring is significant. The KS indicator is also elevated at radial locations affected by a faulty \texttt{XSPEC} measurement (Fig.~\ref{fig:result_100ks_z2_grouped_observables}, right, at $R \simeq R_{500}/2$ in this example).
Such a systematic error is not solely due to poor statistics, nor to inhomogeneity in the cluster gas distribution. Indeed, we have shown that this error acts as a floor uncertainty, inherent to our analysis setup.
First, the inability of a single APEC model to account for a multi-temperature plasma projected along the line-of-sight induces discrepancies that are not well captured by the \texttt{XSPEC} error bars. Development of multi-temperature spectral models with arbitrary distributions of emission measure (e.g., generalizing the class of \texttt{gadem} models) would certainly benefit high-resolution spectroscopy of diffuse astrophysical plasmas. Moreover, identification of point sources contaminating the spectral measurements and buried in the cluster emission should also enhance the quality of spectral fits.
Second, we have worked under the assumption that emission-measure weighting fairly represents the measured X-IFU spectroscopic temperature. Previous studies focusing on \emph{XMM}-Newton and \emph{Chandra} have instead proposed `spectroscopic-like' weightings in order to alleviate this concern \citep[e.g.][]{Mazzotta2004, Vikhlinin2006b}. We have verified that spectroscopic-like temperature profiles are even more discrepant with measurements than emission-measure weighted profiles. Such a work is pending realization in the case of high-resolution instruments like the X-IFU.
More generally, our study calls for further development of new analysis tools dedicated to the analysis of hyperspectral imaging of extended structures \citep[e.g.][]{Picquenot2019}, with an ability to handle the regime of low number of counts.

\subsection{Impact of deeper observations and of targeting a cluster in a more mature evolutionary stage}

Up to now, our results and discussion have focused on a single galaxy cluster extracted from the $z=2$ simulation snapshot. Observing this cluster of galaxies at later times (i.e.,~at lower redshifts, $z=1$ and $1.5$) brings a supplementary amount of information to our study. At later epochs, this cluster is more massive, more extended, hotter and intrinsically more luminous (Table~\ref{table:ce22}). This leads to an increase of the signal-to-noise ratios in observables, both the surface brightness and spectra. Being closer to the observer, the surface brightness is also less faint, hence another increase in signal-to-noise ratios. We note however that the angular diameter distance hardly changes over this range of redshifts, by a few percent at most. Therefore, little gain is expected from angular resolution effects. We have replicated the analysis shown previously for all 14 configurations displayed in Table~\ref{table:KS}. In each case we inferred the four thermodynamic profiles and the three abundance profiles, accounting for the intrinsic dispersion within a radial shell. We inspected the results in light of deviations from the known input profiles. Although the fidelity of the profile reconstruction has a strong radial dependence, we summarise here our results with a single quantity, namely the average of the KS tests performed over the whole range of radii comprised between the PSF size and $R_{500}$ (Table~\ref{table:KS}).

In general, moving the cluster closer to the observer provides an enhancement in the accuracy of the reconstructed thermodynamic profiles and abundance profiles, as does increasing the exposure time. This is especially visible for the density profiles, whose inference relies primarily on surface brightness profiles.
However, we found significant outliers to this overall trend, due to the systematic effects already discussed in Sect.~\ref{sect:results} and App.~\ref{app:result_z2_1Ms}. Indeed, a single faulty temperature measurement in one radial bin (e.g.,~due to mixing components by projection along the line-of-sight) has a negative impact on all reconstructed quantities. Surprisingly, such situations may occur even at low redshifts and/or for large exposure times. One such example is the $z=1$, `irregular' configuration, which seems better characterised at $100$\,ks than at $250$\,ks. A second example is the $z=2$, $250$\,ks, `regular' configuration, which comprises a catastrophic temperature measurement, hence shifting the reconstructed profiles considerably away from the true one.
We also find that observing the cluster in an orientation that minimises the projection of gas phases along the line-of-sight and maximises the asymmetries in the plane of the image (i.e.,~the so-called `irregular' case) often slightly improves the reconstruction of profiles, consistent with our expectations.

    \subsection{Perspectives}

This work presents results on one single test case only. This object may display peculiarities that are not representative of the entire population of groups and clusters. A complete assessment of the scientific feasibility related to thermodynamic profiles inference with X-IFU would involve a larger sample of objects. On the one hand, singularities associated to a single test case would average out~; on the other hand, this would more closely match the approach that observers take in studying intra-cluster and intra-group physics.

The study presented in this paper was conducted with the current public science requirements for the \emph{Athena} missions and the X-IFU instruments. The \emph{Athena} mission is currently undergoing a complete reformulation of its science case and consequently of the specifications of its instruments. The outcomes of our study might be modulated by the outcome of this reformulation. 

As stressed in the introduction, we limited our study to an investigation with the X-IFU instrument following the specifications from the Athena Mock Observing Plan. However, we note that a natural extension of the presented work would be to investigate distant groups of galaxies with deep pointed observations with the second Athena instrument, the Wide Field Imager \citep[WFI,][]{rau17}, and in combination with X-IFU observations. This would optimise the physical characterisation of these objects, at the expense of exposure time as the two instruments will not be observing simultaneously.

Accounting for the current Athena mock observing plan specifications and the ongoing reformulation process for the Athena mission, we will implement this dual combination in a forthcoming investigation.
The upcoming XRISM mission will soon provide the community with unprecedented X-ray observations with high spectral resolution of nearby bright objects. More distant objects such as the first groups of galaxies will have to wait for the advent of observations by the next generation of X-ray integral field unit such as the X-IFU instrument on board the \emph{Athena} mission or the LEM mission concept \citep{Kraft2022}.






\begin{acknowledgements}
The authors would like to thank Dominique Eckert for refereeing this paper and for providing insightful comments on the study. EP, FC and NC acknowledge the support of CNRS/INSU and CNES.
JS acknowledges the support of The Netherlands Organisation for Scientific Research (NWO) through research programme Athena 184.034.002. YMB gratefully acknowledges funding from the Netherlands Research Organisation (NWO) through Veni grant number 639.041.751, and financial support from the Swiss National Science Foundation (SNSF) under project 200021\_213076.

\end{acknowledgements}

%
%

\bibliographystyle{aa} 
\bibliography{biblio} 

\appendix

\section{Results of our validation tests}

\begin{figure*}
\begin{center}
    \begin{tabular}{cc}
    \includegraphics[width=0.48\linewidth]{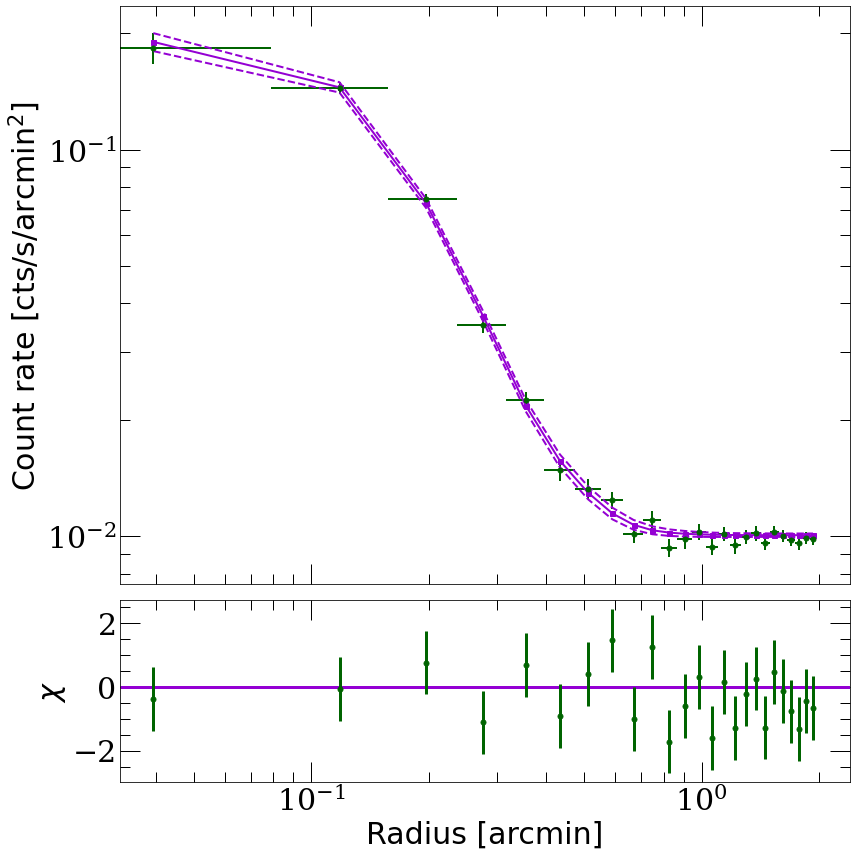} &
    \includegraphics[width=0.48\linewidth]{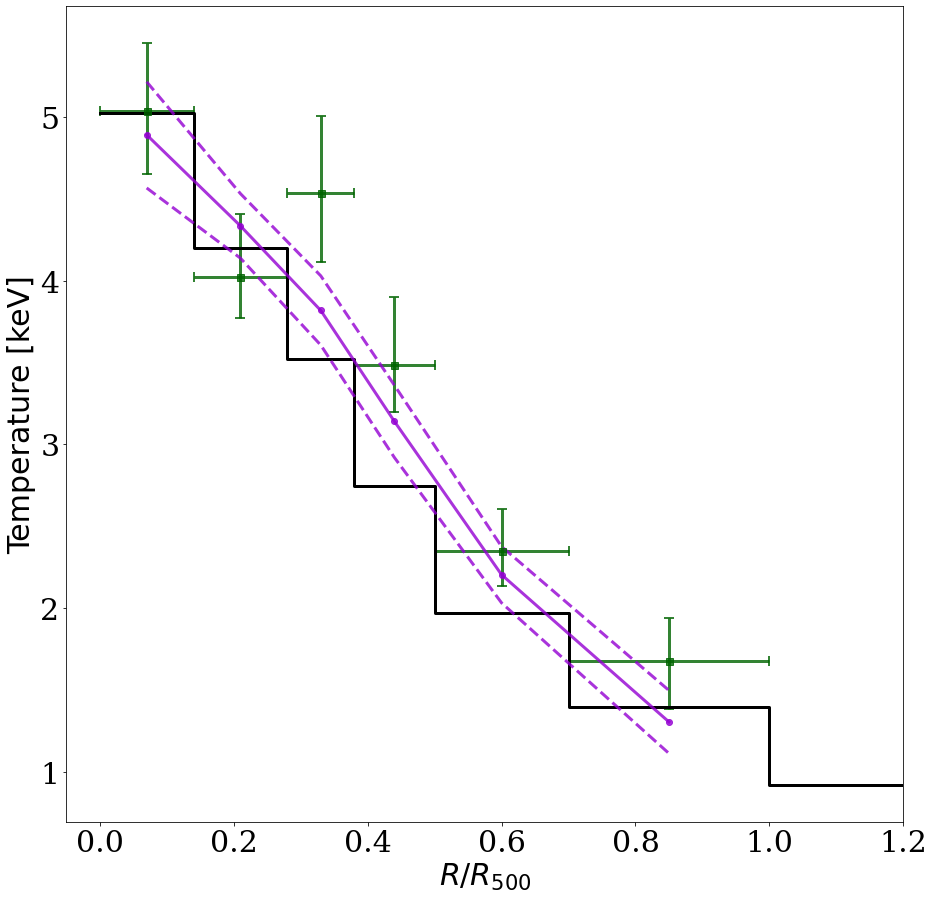} \\
    \includegraphics[width=0.48\linewidth]{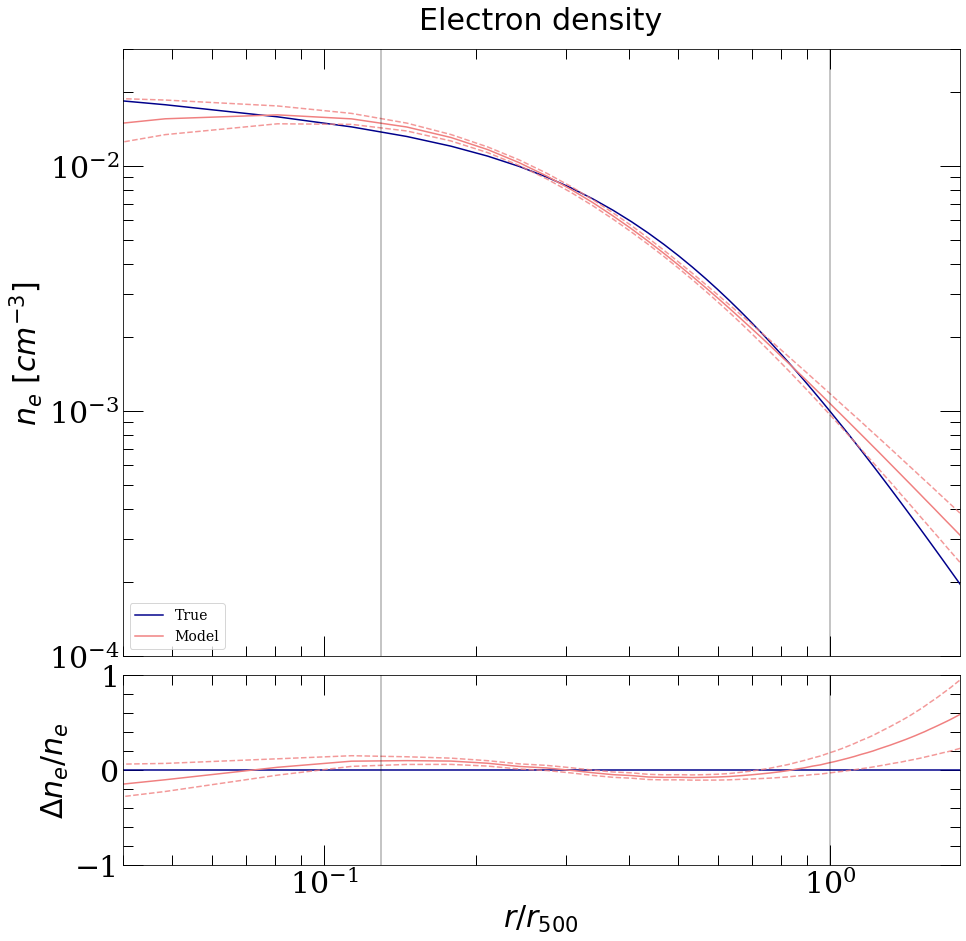} &
    \includegraphics[width=0.48\linewidth]{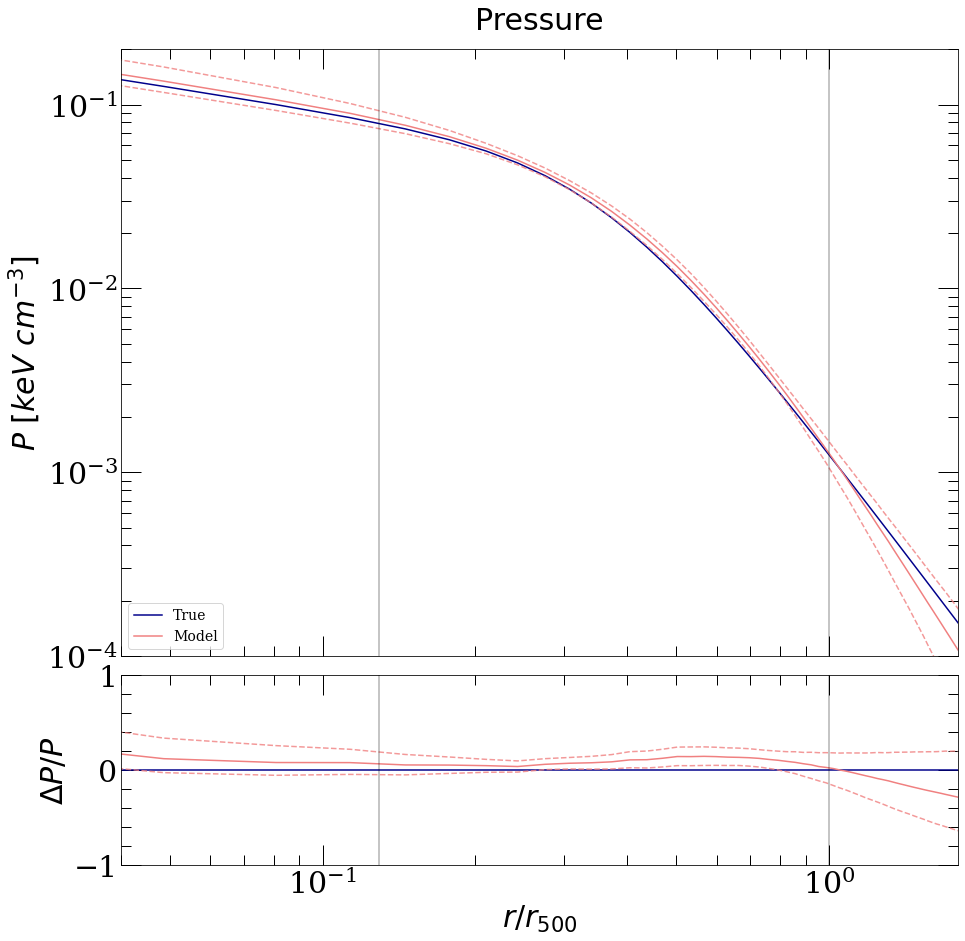} \\
    \end{tabular}
\caption{Validation of the forward-modelling procedure. The input model is a spherically symmetric cluster at $z=2$, transformed into a 100\,ks mock X-IFU observation.
Top left: surface brightness profile from the mock 0.4--1\,keV image (green points and errors). The posterior mean (`best-fit model') and 68\% confidence range appear as a purple solid line and dashed envelope. The sub-panel below represents the difference between measurements and the best-fit model, normalised by the uncertainties.
Top right: two-dimensional temperature profile measured from X-IFU mock events using \texttt{XSPEC} (green points and errors). The emission-measure weighted temperature $T_{EM}$ known as input is displayed with a thick black line. The posterior mean (`best-fit model') and 68\% envelope are shown in purple with solid and dashed lines respectively.
Bottom panels: the inferred electron density (left) and pressure (right) profiles are represented with the red line and dashed envelope (68\% confidence level). The input profiles are displayed in blue and follow Eq.~\ref{eq:density_profile} and~\ref{eq:pressure_profile} respectively. Both sub-panels below represent the deviation of the best-fit models relative to the true input profiles. The range of fidelity of our fit is located between the PSF size (indicated with the leftmost vertical dashed line) and $R_{500}$ (rightmost line).
\label{fig:validation}
}
\end{center}
\end{figure*}

\begin{figure}
\begin{center}
    \includegraphics[width=\linewidth]{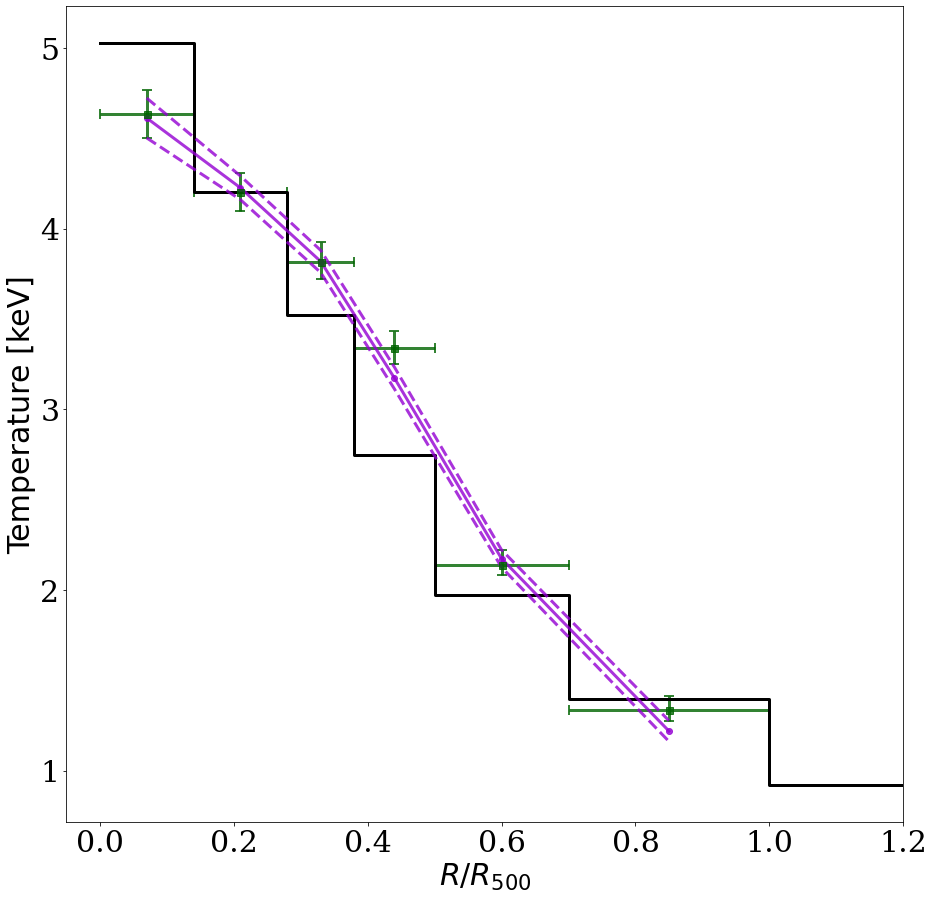}
\caption{Similar figure as Fig.~\ref{fig:validation} (top-right), but for a 1\,Ms X-IFU exposure. This represents the outcome of our validation test in a regime of high photon statistics and it exacerbates the issue due to mixing gas temperatures along the line of sight of this $z=2$ idealised cluster model.
\label{fig:validation_1Ms}
}
\end{center}
\end{figure}

We show here the results of our validation tests on a simpler, spherically symmetric model. These tests are described in Sect.~\ref{sect:validation}. Figures~\ref{fig:validation} and~\ref{fig:validation_1Ms} differ only by the net exposure time of the mock X-IFU observation, respectively 100\,ks and 1\,Ms.

\begin{figure}
\begin{center}
    \includegraphics[width=\linewidth]{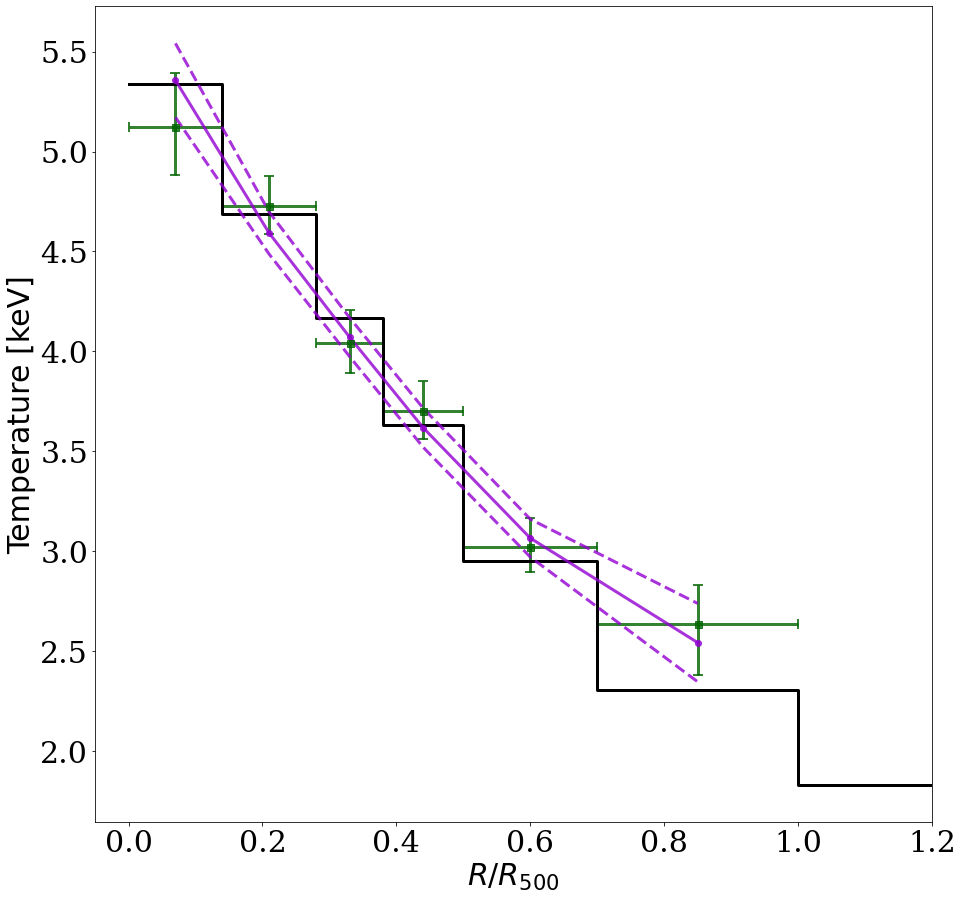}
\caption{Similar figure as Fig.~\ref{fig:validation} (top-right), but for a validation cluster placed at $z=1$ (still at a 100\,ks X-IFU exposure). Because each resolution element in the X-IFU image sees a smaller region (in units of $R_{500}$), mixing effects appear less prominent than in the $z=2$ case~; the reconstructed and \texttt{XSPEC}-fitted temperature profiles are therefore closer to the input EM-weighted profile.
\label{fig:validation_z1}
}
\end{center}
\end{figure}

Fig.~\ref{fig:validation} is related to our baseline validation at 100\,ks X-IFU exposure for a $z=2$ cluster. Fig.~\ref{fig:validation_1Ms} refers to the same cluster, but for a 1\,Ms exposure and we highlight residual uncertainties not solved by increasing the photon statistics. Finally, Fig.~\ref{fig:validation_z1} is for a 100\,ks exposure of a much nearer $z=1$ prototypical cluster, whose physical and angular sizes ($R_{500}$ in kpc and in arcminutes) are about twice that of the $z=2$ cluster. In this case the cluster emission is better resolved by the X-IFU instrument, and mixing effects in the spectral fits are less of a concern.

\section{Results for a 1\,Ms exposure on the $z=2$ cluster}
\label{app:result_z2_1Ms}

\begin{figure}
\begin{center}
\includegraphics[width=\linewidth]{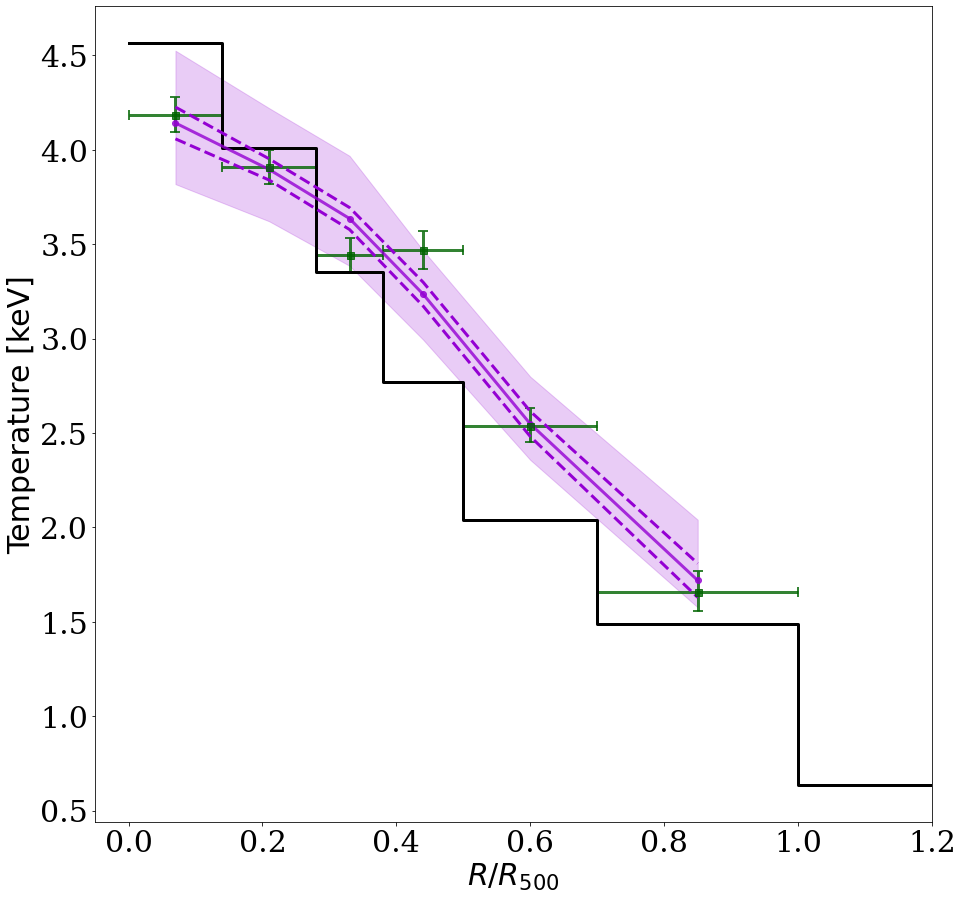}
\caption{Similar figure as Fig.~\ref{fig:result_100ks_z2_grouped_observables} (right), but for a 1\,Ms mock X-IFU exposure. Despite the 10-fold increase in photon statistics, the complex structure of the $z=2$ Hydrangea cluster prevents us from achieving a perfect reconstruction of the projected temperature profile, mainly because of faulty \texttt{XSPEC} single-temperature fits (green points and errors).} 
\label{fig:result_1Ms_z2_grouped_observables}
\end{center}
\end{figure}

\begin{figure}
\begin{center}
\includegraphics[width=\linewidth]{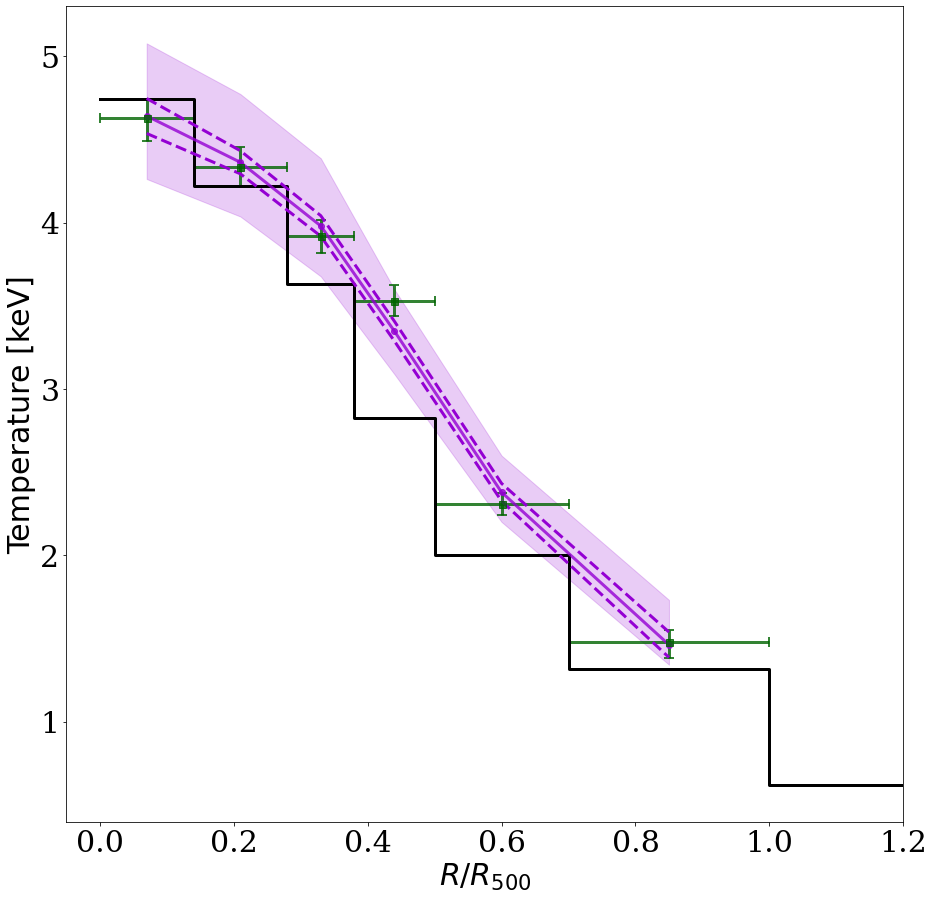}
\caption{Similar figure as Fig.~\ref{fig:result_1Ms_z2_grouped_observables}, but the cluster is observed along a line-of-sight that minimises projection effects. Mixing issues are mitigated by selecting this cluster orientation, enabling a more faithful reconstruction of the cluster profiles.} 
\label{fig:result_1Ms_z2_spread_observables}
\end{center}
\end{figure}

In order to gauge the ultimate capabilities of X-IFU in precisely determining the gas content of the $z=2$ cluster, we reproduced the experiment of Sect.~\ref{sect:results} using a 1\,Ms exposure instead of 100\,ks. We illustrate two salient results obtained by increasing the exposure time by a factor 10. On the one hand, \texttt{XSPEC} temperature measurements show reduced error bars due to higher signal-to-noise ratios in the fitted spectra (Figs.~\ref{fig:result_1Ms_z2_grouped_observables}), however the disagreement with the EM-weighted input model still persists. As hinted in our validation experiment (Sect.~\ref{sect:validation}), this may be attributed to projection effects mixing components in the resulting observables, which are not solved solely by increasing the statistics. We demonstrate this by running our analysis on the same cluster, albeit observed from an alternative orientation selected so that it minimises the projection effects along the line-of-sight. Fig.~\ref{fig:result_1Ms_z2_spread_observables} shows the result obtained for that configuration~; clearly the green, black and purple curves appear closer to each other, despite some residual deviations due to remaining mixing effects.

\begin{figure*}
\begin{center}
\includegraphics[width=0.95\linewidth]{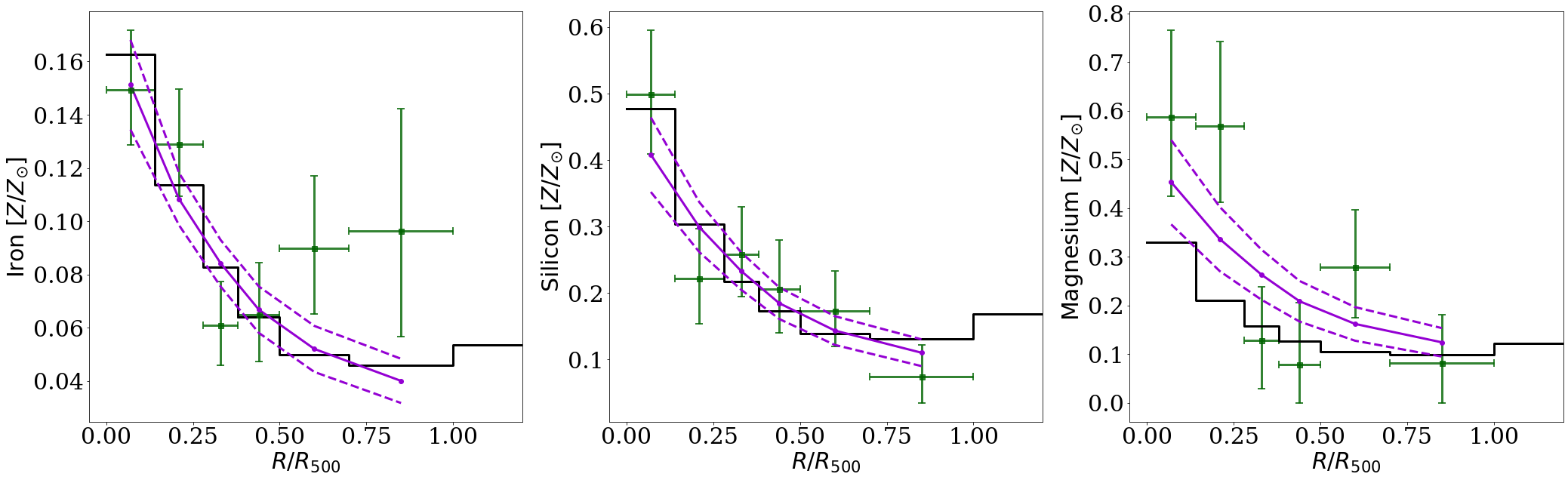}
\caption{Similar figure as Fig.~\ref{fig:result_100ks_z2_grouped_abundances}. but for a 1\,Ms mock X-IFU exposure.}
\label{fig:result_1Ms_z2_grouped_abundances}
\end{center}
\end{figure*}

On the other hand, the reconstruction of abundance profiles at a 1\,Ms exposure (Fig.~\ref{fig:result_1Ms_z2_grouped_abundances}) appears more satisfactory when compared to the 100\,ks case (Fig.~\ref{fig:result_100ks_z2_grouped_abundances}), with both iron (Fe) and silicon (Si) profiles well recovered by the forward model. Only magnesium still presents diverging results despite the 10-fold increase in photon statistics.

\end{document}